# Evidence for ultra-water-rich ammonia hydrates stabilized in icy exoplanetary mantles

Anshuman Mondal[1], Katharina Mohrbach[1,2], Timofey Fedotenko[2], Mandy Bethkenhagen[3], Hanns-Peter Liermann[2] and Carmen Sanchez-Valle[1] *

[1]Institut für Mineralogie, Universität Münster, 48149 Münster, Germany

[2]Deutsches Elektronen-Synchrotron DESY, Notkestrasse 85, 22607 Hamburg, Germany

[3]Ecole Polytechnique-LULI, Av. Fresnel, 91120 Palaiseau, France

* corresponding author: sanchezm@uni-muenster.de

## Abstract

Understanding the behavior of the water-ammonia system at high pressure-high temperature conditions is important for modeling the internal dynamics of exoplanet icy mantles. Conventionally, mixtures of ammonia hemihydrate "AHH" (2:1 ammonia-water molar ratio) and $H_2O$ ice VII have been regarded as the ultimate solid phase assembly in the system. Here we report evidence for chemical reactions between AHH and ice VII above 750 K and 16 GPa that stabilize water-rich ammonia hydrates, including a novel ultra-water rich hydrate $NH_3 \cdot 6H_2O$ (1:6 ratio) coexisting with ammonia dihydrate "ADH" (1:2 ratio) and excess ice VII. This assembly is stable up to at least 30 GPa and 1600 K and can be quenched to room temperature. Our results demonstrate that water-rich ammonia hydrates are favored in the icy mantle of 1-2 $M_{Earth}$ exoplanets regardless of the ammonia content of the hydrate crystallized during accretion and/or evolution as long as excess $H_2O$ ice is available. The buoyancy contrast between water-rich hydrates and ice VII may lead to chemical stratification in exoplanet icy mantles, hence affecting their thermal evolution.

## Introduction

Thousands of sub-Neptunes have been identified to date by missions such as Kepler or TESS (1, 2), among which water-rich exoplanets (3) may be an abundant yet poorly understood class. Water-rich interiors (4–6) are possible scenarios for low-density planets of mass below 10 Earth masses ($M_{Earth}$) that are hosted by stars of very low-mass (5, 7) and/or low content of short-lived radionuclide Aluminium-26 ($^{26}Al$) (8). These planets remain a challenge for planet formation theories (9) because they lack an equivalent in the Solar system, where volatiles consolidate either in the small icy satellites (10) or in the solar ice giants (Uranus and Neptune) (11–13). The formation and evolution of water-rich exoplanets with intermediate masses (1-10 $M_{Earth}$), which can be reconstructed from their present-day interior structure (e.g., (14, 15), thus hold important clues to the origin of both solar and extra-solar systems. While direct measurements of exoplanets atmospheric compositions by ongoing space missions (e.g. JWST, PLATO 2.0 and ARIEL) will provide key constraints to interior structure models (16, 17), current inferences largely rely on the interpretation of geophysical observables. This includes, for instance, the mass-radius (M-R) (3, 6, 18, 19) relations or the gravity field data, which is parameterized by the Love numbers (20–22) to describe the internal density distribution. The increased accuracy of the geophysical observations must thus be matched by improved interior models supported by a better description of the planet building blocks, such as ice and silicates, which are often not available.

Interior models for water-rich exoplanets often assume differentiated layers including a small rocky mantle/core underlying ice layers capped by a surface ocean and/or enveloped by a thick helium/hydrogen atmosphere (e.g., (14, 15, 18)). The ice layers, which can extend over 3000-6000 km for exoplanets with 1-5 $M_{Earth}$ (18), are thought to consist of a mixture of water ($H_2O$), ammonia ($NH_3$) and methane ($CH_4$) ices (15), along with salt impurities (23) and volatiles like hydrogen or helium (14). The extreme pressure (P) - temperature (T) conditions in exoplanet interiors (18) stabilize unusual H-C-N-O compounds (24–27), some of which display exotic transitions from molecular to ionic or superionic states (28–33). At present, however, it remains unknown how the different ice phases are distributed throughout the icy mantles, i.e., whether they occur as homogeneous mixtures reflecting the overall composition ratio or as stratified layers of distinct composition due to the buoyancy contrast between constituents. These two possible scenarios would have a distinct impact on the dynamics of the ice layers, which could in turn affect their thermal evolution (34, 35) and the generation of magnetic fields (36, 37) as shown for the ice giant planets. Understanding the phase relations and stability of ice phases at relevant pressure and temperature conditions are thus central to gaining further insights into the interior models of icy planets.

Ammonia-water mixtures are particularly relevant to planetary interiors not only because of their cosmochemical abundance (molar fractions 56% $H_2O$ and 8% $NH_3$) (38) but also due to the range of molecular, ionic and superionic states identified by both experiments (28, 30, 39, 40) and theory (41–43) at conditions prevailing even in the smallest water-rich planets (18). Only three stable ammonia hydrate stoichiometries have been consistently identified in the $NH_3$-$H_2O$ system to date (44): ammonia dihydrate (ADH, $NH_3 \cdot 2H_2O$), ammonia monohydrate (AMH, $NH_3 \cdot H_2O$), and ammonia hemihydrate (AHH, $2NH_3 \cdot H_2O$), each of them displaying extensive polymorphism. Regardless of composition, ammonia hydrates adopt at high pressure a disordered alloy structure characterized by the random distribution of N/O in a bcc lattice and orientational disorder in the protons on all the lattice sites, initially referred to as disordered molecular alloy (DMA) (40, 45); most recent studies report however the partial ionization of the structure, thus renamed as disordered ionico-molecular alloy (DIMA) (28), which may further transform into pure ionic states (39). The possibility of stabilizing one of the ammonia hydrate stoichiometries in planetary interiors may be controlled by the extent of $NH_3$ enrichment achieved during condensation and the specific evolutionary path (11–15). An important role in planetary environments has been assigned to AHH, which appears in coexistence with ice VII upon breakdown/dehydration of the other two $H_2O$-rich stoichiometric phases, AMH and ADH, at low pressure (<10 GPa) and by crystal fractionation upon compression of $H_2O$-rich $NH_3$-$H_2O$ mixtures (40). Therefore, mixtures of AHH and ice VII have been seen as the ultimate solid phase assembly in the $NH_3$-$H_2O$ system (41), hence suggesting that ammonia-rich phases may dominate the compositional structure of water/ice-rich bodies. Yet, the stability of this phase assembly under high pressure-temperature conditions relevant for planetary interiors remains unknown.

Here we investigate the stability and reactivity of AHH and $H_2O$ ice VII mixtures to mantle pressures and temperatures in 1-2 $M_{Earth}$ water-rich exoplanets (18) up to 45 GPa and 1500 K using a multi-technique approach that combines X-ray diffraction (XRD) probes with resistively heated dynamic and laser-heated diamond anvil cells (hereafter RHdDAC and LHDAC, respectively). We systematically mapped the phase relations in the binary system in multiple experimental runs (*SI Appendix,* Fig. S1) that reveal the formation of a previously unknown ultra-water-rich ammonia hydrate above 16 GPa and 750 K. Details about the experimental procedure can be found in the *Materials and Methods*.

## Results and discussion

### *Polymorphism in the AHH-ice VII binary system at high pressure-high temperature conditions*

In all experimental runs exploring the AHH-ice VII binary system (*SI Appendix,* Fig. S1), reflections in the XRD patterns collected before heating and/or compression can be indexed by the monoclinic ($P2_1/c$) AHH-II phase (46, 47), ice VII and the gold (Au) pressure marker (Figs. 1 and 2). Upon heating and/or compression, we have identified a total of five polymorphs of the AHH phase coexisting with ice VII up to 750 K and 45 GPa (Fig. 1 and *SI Appendix* Figs. S2 and S3). Four of these phases are broadly consistent with those reported for the pure AHH system (40, 42, 47, 48) while a previously unknown polymorph is identified above 650 K in the 13-20 GPa pressure range (Fig. 1).

Upon heating monoclinic AHH-II below 500 K in the 4 to 13 GPa pressure range, we observe the replacement of the three main reflections in the 11 to 12 degrees 2θ angle range by a single strong reflection, that suggest the transition towards a higher symmetry AHH phase (*SI Appendix* Fig. 2). Even though only one reflection is well resolved in the pattern due to low scattering efficiency of the ammonia hydrate phase, the observations and pressure-temperature conditions are consistent with the transition towards a body-centered cubic (bcc) phase similar to that reported for the pure AHH system in previous experimental and theoretical studies (39, 40, 42, 48). This phase is distinct from the DIMA phase identified in pure AHH above 20 GPa both at room and high temperature despite the similar bcc structure and N/O disorder (48). This phase is hereafter named as AHH-IV. (Note: An AHH polymorph stable above 69 GPa at room temperature has been reported in the pure AHH system and already named AHH-III (39) and hence the choice of notation). The AHH-IV + ice VII phase assembly appears to be stable up to 700 K below 10 GPa and after melting, it recrystallizes upon cooling (and pressurization) below 500 K following the sequence: ice VII + $NH_3$-rich liquid → ice VII + AHH-IV (*SI Appendix,* Fig. S1). Interestingly, the AHH-IV transforms back into AHH-II upon quenching, demonstrating the reversibility of the transition (*SI Appendix*, Fig. S2). In the pure AHH system, AHH-IV has been identified as a plastic (and disordered) phase by Raman spectroscopy (48) (labeled as AHH-pbcc (48)) and *ab initio* studies (42). Yet, XRD is not a sensitive technique to probe this behavior and additional studies by Raman spectroscopy in the binary system would be necessary to elucidate whether the plastic behavior of AHH is affected by ice VII.

Compression of AHH-II above 13 GPa reveals the formation of a distinct phase, named AHH-V hereafter, which is stable up to ~650 K in the 13-20 GPa pressure range (Fig. 1a). The XRD patterns suggest higher symmetry compared to the AHH-II phase, consistent with the quasi-bcc structured phase identified in the same pressure range in the pure AHH above 450 K (48). Unlike in the pure AHH phase (48), we observe the AHH-V directly upon compression of the AHH-II phase at 298 K. Additionally, we show that AHH-V can be quenched to room temperature (Fig. 1A). The experimental patterns are not consistent with any of the quasi-bcc structures theoretically predicted (41) and hence, the structure of AHH-V remains unresolved (*SI Appendix,* Text S1 and Fig. S4). Above 650 K, AHH-V undergoes yet another transition to a previously unidentified polymorph, hereafter labeled as AHH-VI (Fig. 1B), and characterized by a complex XRD pattern. Possible structures for the newly identified AHH-VI phase are discussed *SI Appendix,* Text S1. We rule out that this phase results from the reaction of AHH-V with ice VII; the isothermal compression of AHH-VI at 701 K reveals changes in the spectra above 24 GPa - replacement of the reflections in the 10-12 degrees 2θ range by a single reflection at 12.5 degree (*SI Appendix,* Fig. S3) - that are consistent with the transition towards a bcc-structured disordered alloy phase reported in the pure AHH system above 20 GPa between 298 and 700 K (48). The volumes (per molecule), derived for the high pressure-high temperature phase (referred to as AHH-VII) assuming a bcc structure (see Materials and Methods), align well with those available for the disordered alloy phase in pure AHH at similar conditions, e.g., 15.66 cm$^3$/molec. (28.2 GPa, 701 K) vs. 15.34 cm$^3$/molec. (28.2 GPa, 650 K) (48) and 15.49 cm$^3$/molec. (29.3 GPa, 701 K) vs. 15.27 cm$^3$/molec. (29.3 GPa, 650 K) (48), further supporting the phase assignment. The disordered alloy phase in pure AHH has been identified as partially ionic by Raman spectroscopy (48) and hence, it is labeled here as AHH-VII D(I)MA due to the lack of information about the molecular/ionic state of the phase in the present study as discussed below.

In an attempt to precisely constrain the stability field of the AHH-V, AHH-VI and AHH-VII D(I)MA phases, we performed continuous compression experiments (49, 50) in the RHdDAC along isothermal paths at 483 ± 5 K and 701 ± 5 K (*SI Appendix,* Fig. S1). Starting with AHH-II phase (coexisting with ice VII), the compression ramp at 483 ± 5 K brackets the transition to AHH-V at 15 GPa and the subsequent transformation into the AHH-VII D(I)MA phase above 21.5 GPa (*SI Appendix,* Fig. S3A). At 701 ± 5 K, AHH-VI transforms into AHH-VII D(I)MA at 24 GPa, and this phase remains stable in the presence of ice VII up to at least 45 GPa below 750 K (*SI Appendix,* Fig. S3B). We also observe the formation of the AHH-VII D(I)MA phase upon compression of AHH-II above 19.3 GPa at room temperature (Fig. 2). The transition pressure is consistent with the lower bound of pressures reported in the literature (47, 48) for the pure AHH system, suggesting little effect of ice VII on the transition. However, our data up to 750 K and 45 GPa neither allow resolving any broadening of the reflections nor the appearance of additional reflections featuring the distortion of the cubic lattice associated to the partial ionization of the structure (28). Additional Raman spectroscopic analysis, which are more sensitive to the presence of ionic species (39, 48), would thus be required to assess the molecular/ionic character of the AHH-VII phase in the binary system.

## *A novel ultra H₂O-rich ammonia hydrate in the NH₃-H₂O system*

Upon heating the AHH + ice VII assembly above 750-800 K (> 15 GPa), we consistently observe the appearance of new features in the XRD patterns (Fig. 2 and *SI Appendix,* Fig. S5). The characteristic reflections for the starting AHH phase (AHH-V or AHH-VII D(I)MA, *SI Appendix,* Fig. S1) are replaced by two additional reflections at 2θ angles between those of the original AHH phase and ice VII, the latter being always observed in the XRD pattern (see *SI Appendix,* Fig. S6 for details). These changes cannot be ascribed to additional polymorphs or to the formation of a superionic AHH phase because the new reflections are consistent with those appearing at similar conditions in the AMH- and ADH-ice VII systems (*SI Appendix,* Figs. S7 and S8), as discussed later. Instead, we interpret the observations as the result of chemical reactions between the AHH phases and ice VII, which lead to the formation of a distinct phase assembly composed of two ammonia hydrate phases in coexistence with excess ice VII (Fig. 2).

The reaction has been confirmed in three independent runs upon heating AHH-V + ice VII and AHH-VII D(I)MA + ice VII between 750 and 1450 K in the 15 to 25 GPa pressure range, in both RHdDAC and LHDAC experiments (Fig. 2 and *SI Appendix,* Figs. S1 and S5). Two additional LHDAC runs also bracketed the reaction at 18.6 and 30.2 GPa respectively, although the temperature could not be accurately constrained by spectroradiometry (*SI Appendix,* Table S1). Interestingly, the assembly can be quenched to room temperature (Fig. 2), and it is stable upon further compression up to at least 25.1 GPa at room temperature (Fig. 2 and *SI Appendix,* Table S2). The diffraction patterns recorded both at high P-T and upon quenching can be indexed by a combination of two bcc-structured ammonia phases and ice VII (Fig. 2 and *SI Appendix,* Figs. S5 and S6). The reflections from one of the phases -hereafter referred to as 'reacted' ammonia hydrate- are largely consistent with those of the ADH-DMA phase (40, 50), while those of the other phase -hereafter referred to as novel ammonia hydrate- do not match any of the phases identified in the NH₃-H₂O system up to now.

In the absence of single-crystal data, and to gain further insights into the nature and stoichiometry of the novel and 'reacted' ammonia hydrates, we compare the volumes (per molecule), retrieved upon quenching the assembly to room temperature at different pressures (*Materials and Methods* and *SI Appendix,* Table S2), with those from known ammonia hydrate DMA/DIMA phases (40, 47, 50, 51), as well as the NH₃ (52) and H₂O ice VII (53) end-members (Fig. 3). The excellent agreement between the volumes for the 'reacted' ammonia hydrate and those recently reported for ammonia dihydrate (NH₃·2H₂O), ADH-DMA (50), further supports the initial phase indexing. As for the novel ammonia hydrate, we compare the volume (per molecule) with the pressure-volume curves calculated for a range of hypothetical ammonia hydrates $NH_3 \cdot xH_2O$ (x = 4, 5, 6, and 7) assuming linear (ideal) mixing in volume between $NH_3$ and ice VII (Fig. 2 and *SI Appendix,* Fig. S9), an approximation supported at high pressure by both theory (54) and recent experimental results in the ADH ($NH_3 \cdot 2H_2O$) system (50). The latter demonstrated that differences between experimental volumes of the ADH-DMA phase and those predicted for $NH_3 \cdot 2H_2O$ by the linear mixing approximation (LMA) decrease as pressure increase, suggesting ideal behavior above 25 GPa (50). Our calculations show that an ammonia hydrate with $NH_3 \cdot 6H_2O$ provides the best fit to all experimental data, including the high-pressure data where the LMA works better (50).

Although other ammonia hydrates stoichiometries have been theoretically predicted in the $NH_3$-$H_2O$ system, including ammonia quaterhydrate (55) (AQH) $4NH_3 \cdot H_2O$, or $2NH_3 \cdot 3H_2O$ (27), the novel hydrate phase is, to the best of our knowledge, the most $H_2O$-rich ammonia hydrate identified to date. The stability of bcc-structured $H_2O$-rich ammonia hydrates with stoichiometries $NH_3 \cdot 6H_2O$ and $NH_3 \cdot 5H_2O$, including at conditions comparable to the experiments, is also confirmed by complementary *ab initio* density functional theory calculations performed in this study (see *Materials and Methods* for details). The calculated structure for $NH_3 \cdot 6H_2O$ at 20 GPa (0 K) is shown in Figure 3; the calculated XRD diffraction pattern compares well with the reflections assigned to the novel hydrate in the experimental patterns at similar pressures (*SI Appendix,* Fig. S10). Upon heating to 900 K, the analysis of the mean square displacement (MSD) curves further confirms the persistence of the bcc structure in the $H_2O$-rich ammonia hydrates, featuring temporary proton hopping along hydrogen bonds at the investigated conditions (*SI Appendix,* Fig. S11). The formation of the novel ammonia hydrate phase would likely involve the exchange of heavy atoms N/O between the bcc lattices of AHH-VII and ice VII, which has been identified as one of the dominant excitations in this temperature regime (< 1000 K) by previous theoretical studies (42).

The chemical reaction between AHH and ice VII at high pressure-temperature conditions can be thus written as:

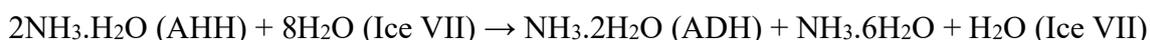

$2NH_3 \cdot H_2O$ (AHH) + $8H_2O$ (Ice VII) → $NH_3 \cdot 2H_2O$ (ADH) + $NH_3 \cdot 6H_2O$ + $H_2O$ (Ice VII)

i.e. the reaction proceeds by the dilution of the AHH that breaks down into two water-rich ammonia hydrates, in presence of excess (unreacted) ice VII, hence showing that the AHH + ice VII is not the ultimate phase assembly in the $NH_3$-$H_2O$ system in contrast to previous reports (39, 41).

## Stability of ammonia hydrates in water-rich environments

The phase diagram for the AHH-ice VII binary system derived from the present work is shown in Fig. 4. Overall, the AHH polymorphs and phase boundaries identified in the binary AHH + ice VII are in good agreement with those experimentally reported for pure AHH (48) in the P-T range of overlap (< 30 GPa and 700 K), with the exception of the new AHH-VI polymorph observed above 650 K between 13-20 GPa. Differences are, however, apparent with the phase diagram of AHH predicted by ab initio simulations (41, 42), which notably fail to resolve the sequence of structural transitions at room temperature and the occurrence of the high temperature phases AHH-IV and/or AHH-VI. We ascribe the disagreement to the limited accuracy of the theoretical structure search due to the reduced number of atoms typically employed in the simulations and the complex H-bonding structure of ammonia hydrates, particularly in the case of disordered phases. Note that inconsistencies between experiments and theory (42, 55) also extend to other ammonia hydrates, and include contrasting results in the phase diagram of AMH (51) or in the stability of the ADH-DMA phase at high pressure (50).

Above 800 K, our results tightly constrain the equilibrium boundary between the AHH + ice VII and the reacted phase assembly between 15 and 25 GPa, which is characterized by a P-T slope of $\delta P/\delta T = 0.159$ GPa/K. Extrapolation of the phase boundary to higher pressures predicts a transition temperature of around 950 K at 45 GPa (Fig. 3), indicating that the assembly ADH-DMA + $NH_3 \cdot 6H_2O$ + $H_2O$ ice VII may be a relevant component in the ice mantle layers of most water-rich exoplanets (56) of the sub-Neptune class.

Further support for this conclusion is provided by the results of similar experiments conducted in the binaries between ice VII and the other known stoichiometric ammonia hydrates, AMH and ADH (Materials and Methods, *SI Appendix,* Figs. S7 and S8). Both AMH and ADH react with $H_2O$ ice VII at high temperature to stabilize the same phase assembly observed in the AHH-ice VII system, ADH-DMA + $NH_3 \cdot 6H_2O$ + $H_2O$ ice VII, as confirmed by the matching reflections in the XRD patterns and the volumes of the phases retrieved upon quenching to room temperature (Fig. 3 and *SI Appendix,* Fig. S9). The combined experiments in the AMH-ice VII constrain the reaction temperature to 800 K – comparable to the AHH-ice VII system – and confirm that it is stable up to at least 31 GPa and 1607 ± 160 K (*SI Appendix,* Fig. S7). While a lower bound for the reaction could not be determined for the ADH-ice VII system here, the reaction is confirmed at 21 GPa and 1582 ± 150 K (*SI Appendix,* Fig. S8). In the absence of excess ice VII, however, we confirm that the ADH-DMA phase remains stable up to at least 21

GPa and 1840 ± 180 K, without breakdown or dehydration into AHH + ice VII as observed at low pressure below 10 GPa (40). These results thus suggest that in the presence of excess ice VII, which is a plausible scenario for water-rich exoplanets, $NH_3$-rich ammonia hydrates AHH and AMH are not stable above 800 K and 16 GPa. Instead, $H_2O$-rich ammonia hydrates appear to be favored at high pressure-temperature conditions, including the newly observed ultra $H_2O$-rich phase $NH_3.6H_2O$, regardless of the ammonia hydrate crystallized during accretion and/or planetary evolution.

## Implications for water-rich exoplanetary interiors

The findings reported here have important implications for the internal structure and dynamics of icy exoplanets, and for the interpretation of remote observables. Among the diversity of size-masses of water/ice-rich exoplanets identified to date (1–7, 18, 22), and their prevailing pressure-temperature conditions, our results directly apply to the smallest exoplanets with ca. 1-2 $M_{Earth}$ displaying high bulk water contents of ca. 50 wt% $H_2O$. The pressure conditions encountered in the ice layer of such planets are expected to be within our experimental pressure range (<30 GPa) at sub-solidus conditions (18). Hence, the water-rich ammonia hydrate assembly -ADH-DMA + $NH_3.6H_2O$ + ice VII- would dominate the compositional structure of a large fraction of their icy mantle provided that temperatures above 800 K are reached and regardless of the ammonia hydrate – ice VII assembly that dominates shallower depth (Fig. 4 and *SI Appendix,* Figs. S7 and S8). In warm TRAPPIST(1f)-like planets (1-1.4 $M_{Earth}$) (7), for instance, the water-rich ammonia hydrate assembly would be stable from 1800 km depth down to the ice/rock interface located at ~2500 km depth (25 GPa (56), Fig. 4).

The shift in the stable ammonia hydrate-ice mixture at mid-mantle depths in 1-2 $M_{Earth}$ exoplanets may encompass changes in the physical properties of the icy layer that, in turn, could affect its dynamics and thermal evolution (14, 15, 18, 20), and may result in anomalies in geophysical observables such the luminosity (57) and the magnetic activity (15, 36, 37). To this extend, water-rich ammonia hydrates, including the novel phase are a potential source for the development of instabilities driven by the density contrast with ice VII. Both ADH and $NH_3.xH_2O$ display densities that are 13% and 6% lower compared to ice VII at 22.3 GPa (Fig. 3), which could promote the formation of ammonia-bearing upwellings within the icy mantle layer. Because convection in the icy layer may be hindered, the survival of the compositional instabilities would result in an efficient mechanism to segregate ammonia-bearing phases from ice VII, which will accumulate at the bottom of the icy mantle forming a differentiated basal layer (Fig. 5). Furthermore, compositional stratification in the icy layer, with ammonia-enriched domains at shallower depths embedded in an ice VII matrix, is expected to affect its rheological properties and thus, the style of convection. Because no strength or viscosity measurements in ice VII or ammonia hydrates at relevant conditions have been reported yet, a quantitative analysis of convection that includes the complex rheology of planetary ice mixtures is not plausible at present. Nevertheless, ADH is several orders of magnitude less viscous than high-pressure $H_2O$ ices, including V, VI and VII, at low pressure conditions and near planetary strain rates (58, 59). If the viscosity contrasts persist at relevant conditions, rheologically weaker ammonia hydrates would act as strain localizing centers within the viscous ice VII (60, 61), which remain essentially undeformed, likely disrupting large-scale convection. Moreover, another important effect of chemical stratification is that the ammonia hydrates domains may act as thermal insulators (62), hence reducing the efficiency of heat transfer and planetary cooling (63, 64) and possibly affecting its luminosity (57).

Finally, the relevance of these results for larger icy bodies would depend on the high pressure-temperature stability of the water-rich ammonia hydrate assembly, which remains unconstrained thus far (Fig. 4). If its stability is limited to 30 GPa - the highest pressure investigated here -, the stratified layer would represent less than 10% of the icy mantle thickness of water-rich exoplanets with more than $5M_{Earth}$ (18) such as GJ 1214b with $6.6M_{Earth}$ (14, 65) or K2-240 with $5M_{Earth}$ (66). While the effect on the dynamics of the global icy mantle may be limited due to the reduced relative thickness of the layer, it may still act as a thermal boundary layer affecting secular cooling (57, 63). In contrast, the transition towards superionic states of the ammonia hydrates first and of $H_2O$ ice at a later stage is expected (30, 32, 33), if the assembly remains stable to higher pressures and temperatures than those explored here (Fig. 4). The presence of superionic phases, together with the stratification of the ice layer, may thus be a potential source for unusual magnetic activity in icy exoplanets as proposed for the ice giants Uranus and Neptune (67, 68).

In conclusion, we report evidence for a novel water-rich ammonia hydrate in the $NH_3$-$H_2O$ system, $NH_3 \cdot 6H_2O$, which is stable below 1800 km depth in icy mantles of TRAPPIST(1f)-like planets. This phase has the highest water content among all the ammonia hydrates identified to date, suggesting a larger diversity of compounds in the canonical $NH_3$-$H_2O$ system at high pressure-high temperature conditions. We show that water-rich ammonia hydrates, rather than ammonia-rich hydrates, are favored in the interior of small water-rich exoplanets (1-2 $M_{Earth}$): this will result in buoyancy-driven chemical stratification in the ice layer that could, in turn, affect the dynamics and thermal evolution of the planet. Future studies are needed to estimate the stability of water-rich ammonia hydrate assemblages, and their transport properties (namely viscosity and thermal conductivity) at conditions comparable to the mantle ice layers of water-rich exoplanets to improve our understanding of their interior structures and evolution.

## Materials and Methods

**Sample preparation.** All investigated samples (AHH + ice VII, AMH + ice VII, ADH + ice VII and ADH-DMA) were prepared from the same commercial $NH_3$-$H_2O$ solution containing 32 wt% $NH_3$ (Sigma-Aldrich, product No. 1.05426) by following different pressure-temperature paths and compression rates in the DAC. Because the concentration of the solution is close to the room temperature solubility of $NH_3$ in $H_2O$, a fresh bottle of solution refrigerated to ~278 K was employed in each loading to prevent $NH_3$ evaporation during the loading procedure. The DAC assembly was also precooled to ~292 K before the loading. The solution was subsequently loaded with the help of a medical syringe and immediately sealed at a pressure below 1 GPa to remain in the liquid state.

Mixtures of AHH (II phase) and Ice VII were synthetized upon slow compression of the solution to 4 GPa at room temperature while the rapid compression of the solution above 7 GPa leads to the formation of the AMH-DMA structure in the presence of excess $H_2O$ ice VII (40). Furthermore, the ADH-DMA samples were prepared by the cryogenic compression procedure described in detail elsewhere (50, 69, 70). Briefly, the refrigerated solution was locked in the DAC, cooled down to approximately 160 K in a liquid $N_2$ bath, and further pressurized to 10 GPa to stabilize the ADH-DMA phase before warming up the cell to room temperature. Some of the attempts resulted in the formation of ADH-DMA coexisting with excess $H_2O$ ice VII (71) due to evaporation of some $NH_3$ during the loading procedure. All samples were inspected by synchrotron powder XRD (see details below) prior to the experiments to verify the phases and the quality of the diffraction patterns. The AHH-II phase displayed a textured powder pattern, whereas both AMH-DMA and ice VII showed quasi-single crystalline pattern consisting of two to three larger spots observed in the Debye rings. The synthesis of ADH-DMA and ADH-DMA + ice VII samples through the cryogenic compression method resulted in excellent quality powder diffraction patterns.

**High temperature diamond anvil cell experiments.** Both resistively heated dynamic diamond anvil cells (RHdDAC) and laser heated diamond anvil cells (LHDAC) were employed to reach temperatures above 1500 K at pressures of up to 45 GPa. All DACs were of symmetric type equipped with 200-300 μm diameter culet diamonds (type-Ia) and X-ray transparent cubic boron nitride (cBN) seats on the detector side (downstream side) to maximize the accessible 2θ angle during the diffraction analysis. Rhenium (Re) gaskets pre-indented to approximately 30-40 μm thickness and lined with a thin gold inlay that served as sample chambers to prevent chemical reactions of the gasket with the sample at high temperature.

RHdDAC experiments were performed using the setup described in Méndez et al. (49), which is based on a modified symmetric-type DAC equipped with two rigid graphite heaters around the diamond anvils and pressurized by a water-cooled piezoelectric actuator. The cell assembly operates in a vacuum vessel to ensure minimum heat loss and to prevent the oxidation of the diamonds and other parts of the cell at high temperature. Additional details about the setup can be found elsewhere (49). Pressure was determined from the thermal equation of state of finely grained gold (72) added to the compression chamber. Estimated errors in pressure are 6% below 600 K and 10% above this temperature. Temperature was monitored to ±5 K by a Rh-Pt alloy R-type thermocouple attached to the diamond culet pavilion on the downstream side of the cell (cBN seat). Note that the use of second thermocouple on the upstream side of the cell (tungsten carbide WC seat) during the experiments was not compatible with the low temperature sample loading procedure described above. Yet, the configuration adopted during the experiments (i.e. thermocouple on the cBN seat side) provides the most accurate estimate of sample temperature as shown in a series of temperature calibration runs conducted before the experiments. For the calibration runs, temperature in the compression chamber was determined from the thermal expansion of Au (73) at ambient pressure measured by X-ray diffraction [Au(111) reflection] and compared to the thermocouple reading. A second thermocouple was attached to the upstream side of the cell (tungsten carbide WC seat) for comparison. Sample temperature was systematically lower than the thermocouples readings, although the closest agreement was found for the thermocouple mounted on the cBN seat side (*SI Appendix,* Fig. S12) due to its higher thermal conduction compared to WC (74). The mismatch between the sample temperature and the thermocouple temperatures increases linearly with temperature and reaches up to 80 K at 900-950 K, the highest experimental temperature in the present study, for the thermocouple on the cBN seat side. A linear fit on the difference in temperatures vs. cBN thermocouple data up to 950 K allows the retrieval of the temperature correction for the subsequent RHdDAC experimental runs (*SI Appendix,* Fig. S12). All temperatures reported in this work are corrected, and hence, they refer to the actual sample temperature encountered in the sample chamber. The reported temperature error corresponds to the intrinsic accuracy of the thermocouple, which is ±5 K.

LHDAC experiments employed the double-sided laser heating setup available at P02.2, the Extreme Conditions Beamline (ECB) of PETRA III at Deutsches Elektronen-Synchrotron, DESY(75). The setup includes a 200 W near-IR Yb-fiber laser, split into two beams, focused into a spot size of approximately 20 μm and a Shamrock spectrometer (SR-303i-A-SIL) to analyze the thermal radiation from both sides of the sample. The transfer function of the optical system, required to separate the external background radiation from the thermal radiation of the sample, was calibrated using a tungsten halogen lamp (OPTEEMA Engineering GmbH, OL-245M-K3) at three different temperatures: 2200 K, 2500 K and 2900 K. Furthermore, background corrected spectra from the hot sample were fit to the ideal Planck's law to extract the sample temperature using the T-Rax software (https://github.com/CPrescher/TRax) in the wavelength range of 450-800 nm. Error in temperatures are estimated to be ~10% of the absolute temperature value obtained from the Plank's fit (76). Single crystal boron-doped diamond disks of 10 μm thickness (50 μm diameter, Applied Diamond, Inc.) embedded into the sample served as laser couplers to ensure efficient heating of the low-Z transparent samples investigated here (*SI Appendix,* Fig. S13). Both the sample and the coupler were thermally insulated from the diamond culet surface by using either alumina ($Al_2O_3$) deposited by chemical vapor deposition CVD (100 nm thick) on the diamond culet or a pellet of silica glass (5 μm thick). During measurements, the sample was compressed using a gas-driven membrane attached to the upstream side of the DAC. Cold sample pressures are reported as the average pressure determined from the equation of state (EoS) of $H_2O$ ice VII (53) and ammonia hydrate D(I)MA phases (50, 51), uncertainties are reported as the standard deviation of the pressure determined from all the calibrants, which does not exceed 1 GPa.

**X-ray diffraction data collection and analysis.** Monochromatic synchrotron X-rays with energies of either 25.6 keV (RHdDAC) or 42.7 keV (LHDAC) were employed to perform XRD experiments at beamline P02.2, DESY (77). The x-ray beam was focused using Compound Refractive Lenses (CRLs) into a spot size of 3 μm x 8 μm for RHdDAC experiments while Kirkpatrick-Baez (KB) mirrors were employed to achieve a beam spot of 2 μm x 2 μm (H x V) in the LHDAC. XRD data were collected using a flat panel detector XRD1621 from Perkin Elmer equipped with a Ta-doped CsI sensor bonded to an amorphous silicon readout chip. Sample to detector distance, tilt and rotation of the detector were calibrated using a $CeO_2$ standard (NIST 674b) and the DIOPTAS software (78).

Experiments in the AHH-ice VII were conducted along 9 different experimental runs both in RHdDAC and LHDAC up to 45 GPa and 1425±140 K. Data were typically collected upon increasing/decreasing temperature following pressure-temperature pathways that allowed tightly constraining multiple phase boundaries in the system (*SI Appendix,* Fig. S1). Two continuous compression experiments along isotherms at 483 ± 5 and 700 ± 5 K were conducted in the RHdDAC setup up to 35 and 45 GPa, respectively, with compression rates of 0.05 GPa/sec. In the LHDAC runs, the XRD and thermal radiation data were collected simultaneously. Typical acquisition times were 5-10 sec per diffractogram for both RHdDAC and LHDAC experiments. Moreover, systematic grid scans were performed across the entire sample chamber for RHdDAC runs and on the hot spot for LHDAC runs, before and after the experiments, and in situ at the highest temperature reached in the RHdDAC runs with exposure times of 1-3 sec per spectra. Data in the AMH-ice VII system was recorded in two experiments in both RHdDAC and LHDAC above 800 K to search for plausible reactions (*SI Appendix,* Fig. S7), while only LHDAC experiments were conducted in the ADH-ice VII and in the pure ADH systems (*SI Appendix,* Fig. S8). The data collection strategy and acquisition times were similar to those described above for the AHH-ice VII system. The low X-ray scattering efficiency of ammonia hydrate phases resulted in a limited number of reflections unambiguously identified for each phase in the recorded diffractograms (e.g., Figs. 1-2 and *SI Appendix,* Figs. S2-S3 and S5-S8). For the bcc or quasi-bcc phases, for instance, three reflections are observed at most in the RHdDAC runs [(110), (200) and (211), Fig. 2] while only the most intense (110) reflection was observed in the LHDAC runs due to the higher background arising from the $Al_2O_3$ or $SiO_2$ thermal insulation layer loaded in the cell (Fig. 2 and *SI Appendix,* Figs. S7 and S8).

Diffraction images were integrated into 1D patterns using the DIOPTAS software (78). For ramp compressions in RHdDAC runs, a quick visualization of all diffractograms recorded through the entire compression run as an intensity plot was generated using the modified DIOPTAS software integrated into the P02.2 data pipeline (79). The weak and limited number of exploitable reflections associated to high symmetry phases precluded the refinement of the XRD patterns using standard multi-peak fitting routines. Instead, the 2θ-angle of individual reflection was determined based on a pseudo-Voigt fit using a least-squares fitting routine in Python (https://gitlab.desy.de/rachel.husband/peakfit) previously validated in the analysis of ADH compression data in dynamic DAC experiments (50).

Volumes (per molecule) for AHH-VII D(I)MA, the 'reacted hydrate' ADH-DMA and the 'novel' ultra-water rich $NH_3 \cdot xH_2O$ ammonia hydrate, were calculated based on a bcc structure and utilizing the most intense reflection, i.e. (110) reflection, in the diffraction patterns recorded at different pressures upon quenching and further compression at room temperature (Fig. 3 *SI*

*Appendix,* Table S2). Errors in volume in the RHdDAC runs were estimated based on the standard deviation in volumes calculated from the (110), (200) and (211) reflections (Fig. 2, *SI Appendix,* Figs. S5 and S7A). Because only the (110) reflection was consistently observed in the LHDAC runs (Fig. 2C, *SI Appendix,* Figs. S7B and S8), the errors in volume are assumed to be similar to those in the RHdDAC runs.

*Ab initio* **simulations.** We conducted density functional theory molecular dynamic (DFT-MD) simulations to directly test the stability of bcc-structured water-rich ammonia hydrates with stoichiometries $NH_3 \cdot 6H_2O$ and $NH_3 \cdot 5H_2O$, including at the pressure-temperature conditions relevant for the experiments, 19 GPa and 900 K. Note that true structure search calculations would be of limited value here as they will never converge into a bcc structure in the case of asymmetric $NH_3$ and $H_2O$ mixtures due to the limited number of atoms that can be realistically included in the calculations, i.e., typically 2-16 formula units (f.u) vs. 49 f.u. required for 1:6 mixture. The calculations were conducted using the VASP code (80–82). We considered simulations cells with 54 molecules (9 $NH_3$ and 45 $H_2O$) and 64 molecules (9 $NH_3$ and 55 $H_2O$) to approximate the stoichiometries of $NH_3 \cdot 5H_2O$ and $NH_3 \cdot 6H_2O$, respectively. The DFT-MD simulations employed the Perdew–Burke–Ernzerhof (PBE) exchange-correlation functional and the hard Projector-Augmented Wave (PAW) pseudopotentials (83) with an energy cutoff of 1000 eV. The initial cold compression curve calculations employed a k-spacing of 0.5 Å$^{-1}$ and varied the position of the $NH_3$ molecules on bcc lattice sites to optimize the free energy. The subsequent finite-temperature calculations were conducted for the $NH_3 \cdot 5H_2O$ phase to optimize the computation time using the Nosé-Hoover thermostat (84) to control the temperature, the Baldereschi Mean Value point (85) to sample the k-space and timestep size of 0.25 fs. The typical simulation duration after equilibration was 5 ps.

## Data availability

All raw XRD images, integrated 1-dimensional diffraction patterns, calculation files, and instruction files to reproduce all the main figures and supplementary figures are available in the Zenodo repository, which will be available upon request to the corresponding author.

## Code availability

The Python least-squared fitting routine employed in the analysis of the X-ray diffraction patterns is available in open access in Github (https://gitlab.desy.de/rachel.husband/peakfit).

# References

1. W. J. Borucki, *et al.*, Kepler Planet-Detection Mission: Introduction and First Results. *Science* **327**, 977–980 (2010).

2. N. M. Batalha, *et al.*, PLANETARY CANDIDATES OBSERVED BY *KEPLER* . III. ANALYSIS OF THE FIRST 16 MONTHS OF DATA. *ApJS* **204**, 24 (2013).

3. L. Noack, I. Snellen, H. Rauer, Water in Extrasolar Planets and Implications for Habitability. *Space Sci. Rev.* **212**, 877–898 (2017).

4. M. Lozovsky, R. Helled, C. Dorn, J. Venturini, Threshold Radii of Volatile-rich Planets. *ApJ* **866**, 49 (2018).

5. R. Luque, E. Pallé, Density, not radius, separates rocky and water-rich small planets orbiting M dwarf stars. *Science* **377**, 1211–1214 (2022).

6. D. Valencia, D. D. Sasselov, R. J. O'Connell, Radius and Structure Models of the First Super-Earth Planet. *ApJ* **656**, 545–551 (2007).

7. M. Gillon, *et al.*, Seven temperate terrestrial planets around the nearby ultracool dwarf star TRAPPIST-1. *Nature* **542**, 456–460 (2017).

8. T. Lichtenberg, *et al.*, A water budget dichotomy of rocky protoplanets from 26Al-heating. *Nat. Astron.* **3**, 307–313 (2019).

9. S. N. Raymond, A. Morbidelli, "Planet Formation: Key Mechanisms and Global Models" in *Demographics of Exoplanetary Systems*, Astrophysics and Space Science Library., K. Biazzo, V. Bozza, L. Mancini, A. Sozzetti, Eds. (Springer International Publishing, 2022), pp. 3–82.

10. H. Hussmann, C. Sotin, J. I. Lunine, "Interiors and Evolution of Icy Satellites" in *Treatise on Geophysics*, (Elsevier, 2015), pp. 605–635.

11. W. B. Hubbard, J. J. MacFarlane, Structure and evolution of Uranus and Neptune. *J. Geophys. Res.* **85**, 225–234 (1980).

12. J. J. Fortney, N. Nettelmann, The Interior Structure, Composition, and Evolution of Giant Planets. *Space Sci. Rev.* **152**, 423–447 (2010).

13. R. Helled, N. Nettelmann, T. Guillot, Uranus and Neptune: Origin, Evolution and Internal Structure. *Space Sci. Rev.* **216**, 38 (2020).

14. N. Nettelmann, J. J. Fortney, U. Kramm, R. Redmer, THERMAL EVOLUTION AND STRUCTURE MODELS OF THE TRANSITING SUPER-EARTH GJ 1214b. *ApJ* **733**, 2 (2011).

15. B. Yunsheng Tian, S. Stanley, INTERIOR STRUCTURE OF WATER PLANETS: IMPLICATIONS FOR THEIR DYNAMO SOURCE REGIONS. *ApJ* **768**, 156 (2013).

16. H. Rauer, *et al.*, The PLATO 2.0 mission. *Exp. Astron.* **38**, 249–330 (2014).

17. R. Helled, *et al.*, Ariel planetary interiors White Paper. *Exp. Astron.* **53**, 323–356 (2022).

18. C. Sotin, O. Grasset, A. Mocquet, Mass–radius curve for extrasolar Earth-like planets and ocean planets. *Icarus* **191**, 337–351 (2007).

19. L. Zeng, *et al.*, Growth model interpretation of planet size distribution. *Proc. Natl. Acad. Sci. U.S.A.* **116**, 9723–9728 (2019).

20. H. Hellard, *et al.*, Retrieval of the Fluid Love Number $k_2$ in Exoplanetary Transit Curves. *ApJ* **878**, 119 (2019).

21. Sz. Csizmadia, H. Hellard, A. M. S. Smith, An estimate of the $k_2$ Love number of WASP-18Ab from its radial velocity measurements. *A&A* **623**, A45 (2019).

22. G. Tobie, O. Grasset, C. Dumoulin, A. Mocquet, Tidal response of rocky and ice-rich exoplanets. *A&A* **630**, A70 (2019).

23. J.-A. Hernandez, R. Caracas, S. Labrosse, Stability of high-temperature salty ice suggests electrolyte permeability in water-rich exoplanet icy mantles. *Nat. Commun.* **13**, 3303 (2022).

24. A. Hermann, N. W. Ashcroft, R. Hoffmann, High pressure ices. *Proc. Natl. Acad. Sci. U.S.A.* **109**, 745–750 (2012).

25. M. Frost, *et al.*, Diamond precipitation dynamics from hydrocarbons at icy planet interior conditions. *Nat. Astron.* **8**, 174–181 (2024).

26. L. J. Conway, C. J. Pickard, A. Hermann, Rules of formation of H–C–N–O compounds at high pressure and the fates of planetary ices. *Proc. Natl. Acad. Sci. U.S.A.* **118**, e2026360118 (2021).

27. A. S. Naumova, S. V. Lepeshkin, P. V. Bushlanov, A. R. Oganov, Unusual Chemistry of the C–H–N–O System under Pressure and Implications for Giant Planets. *J. Phys. Chem. A* **125**, 3936–3942 (2021).

28. C. Liu, *et al.*, Topologically frustrated ionisation in a water-ammonia ice mixture. *Nat. Commun.* **8**, 1065 (2017).

29. C. Cavazzoni, *et al.*, Superionic and Metallic States of Water and Ammonia at Giant Planet Conditions. *Science* **283**, 44–46 (1999).

30. S. Ninet, F. Datchi, A. M. Saitta, Proton Disorder and Superionicity in Hot Dense Ammonia Ice. *Phys. Rev. Lett.* **108**, 165702 (2012).

31. M. Bethkenhagen, D. Cebulla, R. Redmer, S. Hamel, Superionic Phases of the 1:1 Water–Ammonia Mixture. *J. Phys. Chem. A* **119**, 10582–10588 (2015).

32. V. B. Prakapenka, N. Holtgrewe, S. S. Lobanov, A. F. Goncharov, Structure and properties of two superionic ice phases. *Nat. Phys.* **17**, 1233–1238 (2021).

33. G. Weck, *et al.*, Evidence and Stability Field of fcc Superionic Water Ice Using Static Compression. *Phys. Rev. Lett.* **128**, 165701 (2022).

34. W. D. Hubbard, M. Podolak, D. J. Stevenson, "The interior of Neptune" in (University of Arizona Press, 1995), pp. 109–138.

35. N. Nettelmann, R. Helled, J. J. Fortney, R. Redmer, New indication for a dichotomy in the interior structure of Uranus and Neptune from the application of modified shape and rotation data. *Planet. Space Sci.* **77**, 143–151 (2013).

36. S. Stanley, J. Bloxham, Convective-region geometry as the cause of Uranus' and Neptune's unusual magnetic fields. *Nature* **428**, 151–153 (2004).

37. B. Militzer, Phase separation of planetary ices explains nondipolar magnetic fields of Uranus and Neptune. *Proc. Natl. Acad. Sci. U.S.A.* **121**, e2403981121 (2024).

38. M. Asplund, N. Grevesse, A. J. Sauval, P. Scott, The Chemical Composition of the Sun. *Annu. Rev. Astron. Astrophys.* **47**, 481–522 (2009).

39. W. Xu, *et al.*, Ionic Phases of Ammonia-Rich Hydrate at High Densities. *Phys. Rev. Lett.* **126**, 015702 (2021).

40. C. W. Wilson, *et al.*, On the stability of the disordered molecular alloy phase of ammonia hemihydrate. *J. Chem. Phys.* **142**, 094707 (2015).

41. V. Naden Robinson, Y. Wang, Y. Ma, A. Hermann, Stabilization of ammonia-rich hydrate inside icy planets. *Proc. Natl. Acad. Sci. U.S.A.* **114**, 9003–9008 (2017).

42. V. Naden Robinson, A. Hermann, Plastic and superionic phases in ammonia–water mixtures at high pressures and temperatures. *J. Phys.: Condens. Matter* **32**, 184004 (2020).

43. X. Jiang, X. Wu, Z. Zheng, Y. Huang, J. Zhao, Ionic and superionic phases in ammonia dihydrate $NH_3 \cdot 2H_2O$ under high pressure. *Phys. Rev. B* **95**, 144104 (2017).

44. J. S. Kargel, Ammonia-water volcanism on icy satellites: Phase relations at 1 atmosphere. *Icarus* **100**, 556–574 (1992).

45. J. S. Loveday, R. J. Nelmes, Ammonia Monohydrate VI: A Hydrogen-Bonded Molecular Alloy. *Phys. Rev. Lett.* **83**, 4329–4332 (1999).

46. C. W. Wilson, C. L. Bull, G. Stinton, J. S. Loveday, Pressure-induced dehydration and the structure of ammonia hemihydrate-II. *J. Chem. Phys.* **136**, 094506 (2012).

47. C. Ma, *et al.*, Ammonia molecule rotation of pressure-induced phase transition in ammonia hemihydrates $2NH_3 \cdot H_2O$. *RSC Adv.* **2**, 4920 (2012).

48. L. Andriambariarijaona, *et al.*, High pressure–temperature phase diagram of ammonia hemihydrate. *Phys. Rev. B* **108**, 174102 (2023).

49. A. S. J. Méndez, *et al.*, A resistively-heated dynamic diamond anvil cell (RHdDAC) for fast compression x-ray diffraction experiments at high temperatures. *Rev. Sci. Instrum.* **91**, 073906 (2020).

50. A. Mondal, R. J. Husband, H.-P. Liermann, C. Sanchez-Valle, Equation of state of ammonia dihydrate up to 112 GPa by static and dynamic compression experiments in diamond anvil cells. *Phys. Rev. B* **107**, 224108 (2023).

51. H. Zhang, "Experimental investigation of the phase diagram of ammonia monohydrate at high pressure and temperature," Sorbonne Université. (2021).

52. F. Datchi, *et al.*, Solid ammonia at high pressure: A single-crystal x-ray diffraction study to 123 GPa. *Phys. Rev. B* **73**, 174111 (2006).

53. S. Klotz, *et al.*, Bulk moduli and equations of state of ice VII and ice VIII. *Phys. Rev. B* **95**, 174111 (2017).

54. M. Bethkenhagen, *et al.*, Planetary Ices and the Linear Mixing Approximation. *ApJ* **848**, 67 (2017).

55. V. Naden Robinson, M. Marqués, Y. Wang, Y. Ma, A. Hermann, Novel phases in ammonia-water mixtures under pressure. *J. Chem. Phys.* **149**, 234501 (2018).

56. C. T. Unterborn, S. J. Desch, N. R. Hinkel, A. Lorenzo, Inward migration of the TRAPPIST-1 planets as inferred from their water-rich compositions. *Nat. Astron.* **2**, 297–302 (2018).

57. A. Vazan, R. Helled, Explaining the low luminosity of Uranus: a self-consistent thermal and structural evolution. *A&A* **633**, A50 (2020).

58. W. B. Durham, S. H. Kirby, L. A. Stern, Flow of ices in the ammonia-water system. *J. Geophys. Res.* **98**, 17667–17682 (1993).

59. T. Kubo, *et al.*, Rheology of water ice VII in *Photon Factory Activity Report 2013*, (2014), pp. 31, 152.

60. W. B. Durham, S. H. Kirby, L. A. Stern, Effects of dispersed particulates on the rheology of water ice at planetary conditions. *J. Geophys. Res.* **97**, 20883–20897 (1992).

61. M. L. Rudolph, M. Manga, Effects of anisotropic viscosity and texture development on convection in ice mantles. *J. Geophys. Res.* **117**, 2012JE004166 (2012).

62. W. B. Durham, O. Prieto-Ballesteros, D. L. Goldsby, J. S. Kargel, Rheological and Thermal Properties of Icy Materials. *Space Sci. Rev.* **153**, 273–298 (2010).

63. N. Nettelmann, *et al.*, Uranus evolution models with simple thermal boundary layers. *Icarus* **275**, 107–116 (2016).

64. S. Kamata, *et al.*, Pluto's ocean is capped and insulated by gas hydrates. *Nat. Geosci.* **12**, 407–410 (2019).

65. D. Charbonneau, *et al.*, A super-Earth transiting a nearby low-mass star. *Nature* **462**, 891–894 (2009).

66. E. Díez Alonso, *et al.*, Two planetary systems with transiting Earth-sized and super-Earth planets orbiting late-type dwarf stars. *MNRAS* **480**, L1–L5 (2018).

67. R. Redmer, T. R. Mattsson, N. Nettelmann, M. French, The phase diagram of water and the magnetic fields of Uranus and Neptune. *Icarus* **211**, 798–803 (2011).

68. S. Stanley, J. Bloxham, Numerical dynamo models of Uranus' and Neptune's magnetic fields. *Icarus* **184**, 556–572 (2006).

69. A. D. Fortes, *et al.*, The high-pressure phase diagram of ammonia dihydrate. *High Press. Res.* **27**, 201–212 (2007).

70. J. S. Loveday, R. J. Nelmes, C. L. Bull, H. E. Maynard-Casely, M. Guthrie, Observation of ammonia dihydrate in the AMH-VI structure at room temperature – possible implications for the outer solar system. *High Press. Res.* **29**, 396–404 (2009).

71. O. Mousis, J. Pargamin, O. Grasset, C. Sotin, Experiments in the $NH_3$-$H_2O$ system in the [0, 1 GPa] pressure range - implications for the deep liquid layer of large icy satellites. *Geophys. Res. Lett.* **29** (2002).

72. O. L. Anderson, D. G. Isaak, S. Yamamoto, Anharmonicity and the equation of state for gold. *J. Appl. Phys.* **65**, 1534–1543 (1989).

73. M. G. Pamato, I. G. Wood, D. P. Dobson, S. A. Hunt, L. Vočadlo, The thermal expansion of gold: point defect concentrations and pre-melting in a face-centred cubic metal. *J. Appl. Crystallogr.* **51**, 470–480 (2018).

74. H. Hwang, *et al.*, Graphite resistive heated diamond anvil cell for simultaneous high-pressure and high-temperature diffraction experiments. *Rev. Sci. Instrum.* **94**, 083903 (2023).

75. Z. Konôpková, *et al.*, Laser heating system at the Extreme Conditions Beamline, P02.2, PETRA III. *J. Synchrotron. Rad.* **28**, 1747–1757 (2021).

76. L. R. Benedetti, P. Loubeyre, Temperature gradients, wavelength-dependent emissivity, and accuracy of high and very-high temperatures measured in the laser-heated diamond cell. *High Press. Res.* **24**, 423–445 (2004).

77. H.-P. Liermann, *et al.*, The Extreme Conditions Beamline P02.2 and the Extreme Conditions Science Infrastructure at PETRA III. *J. Synchrotron. Rad.* **22**, 908–924 (2015).

78. C. Prescher, V. B. Prakapenka, *DIOPTAS*: a program for reduction of two-dimensional X-ray diffraction data and data exploration. *High Press. Res.* **35**, 223–230 (2015).

79. M. Karnevskiy, *et al.*, Automated pipeline processing X-ray diffraction data from dynamic compression experiments on the Extreme Conditions Beamline of PETRA III. *J. Appl. Crystallogr.* **57**, 1217–1228 (2024).

80. G. Kresse, J. Hafner, *Ab initio* molecular dynamics for liquid metals. *Phys. Rev. B* **47**, 558–561 (1993).

81. G. Kresse, J. Hafner, *Ab initio* molecular-dynamics simulation of the liquid-metal–amorphous-semiconductor transition in germanium. *Phys. Rev. B* **49**, 14251–14269 (1994).

82. G. Kresse, J. Furthmüller, Efficient iterative schemes for *ab initio* total-energy calculations using a plane-wave basis set. *Phys. Rev. B* **54**, 11169–11186 (1996).

83. G. Kresse, D. Joubert, From ultrasoft pseudopotentials to the projector augmented-wave method. *Phys. Rev. B* **59**, 1758–1775 (1999).

84. S. Nosé, A unified formulation of the constant temperature molecular dynamics methods. *J. Chem. Phys.* **81**, 511–519 (1984).

85. A. Baldereschi, Mean-Value Point in the Brillouin Zone. *Phys. Rev. B* **7**, 5212–5215 (1973).

86. A. Boultif, D. Louër, Powder pattern indexing with the dichotomy method. *J. Appl. Crystallogr.* **37**, 724–731 (2004).

87. J. Laugier, B. Bochu, *LMGP-Suite of Programs for the Interpretation of X-ray Experiments* (ENSP/Laboratoire des Matériaux et du Génie Physique, 2004).

## Acknowledgements

This research was supported by the German Science Foundation (Deutsche Forschungsgemeinschaft, DFG) through Research Unit FOR 2440/2 (Grant SA2585/5-1). We acknowledge DESY (Hamburg, Germany), a member of the Helmholtz Association, for providing access to the experimental facility PETRA III and for provision of beamtime at ECB P02.2 beamline. The computational work was supported by the French computational centers TGCC and CINES through the GENCI project GEN15731 (Grant No. A0170915731) and the North German Supercomputing Alliance (HRN) under Grant No. mvp00026. We acknowledge the scientific exchange and support of the Centre for Molecular Water Science (CMWS). We thank M. Roeper from DESY HIBEF, S. Kulkarni from DESY Nanolab and H. Koppetz-Mitra from the Institute of Mineralogy, University of Münster, for technical support. Dr. Martin French is acknowledged for support with the *ab initio* calculations and for insightful discussions during the project, and Dr. Rachel Husband for sharing the python routine to process the X-ray data.

## Author contribution

CSV and HPL conceptualized the research and raised funding; AM prepared the experiments, analyzed the data and produced the figures; AM, KM, TF, HPL and CSV conducted the experiments; MB performed the ab initio calculations; CSV, AM and HPL interpreted the results and wrote the manuscript with input from all authors.

## Competing interests

The authors declare no competing interests.

## Supplementary Information

Supplementary information is available which includes Text S1, Figures S1 to S13 and Tables S1 and S2.

# List of figures

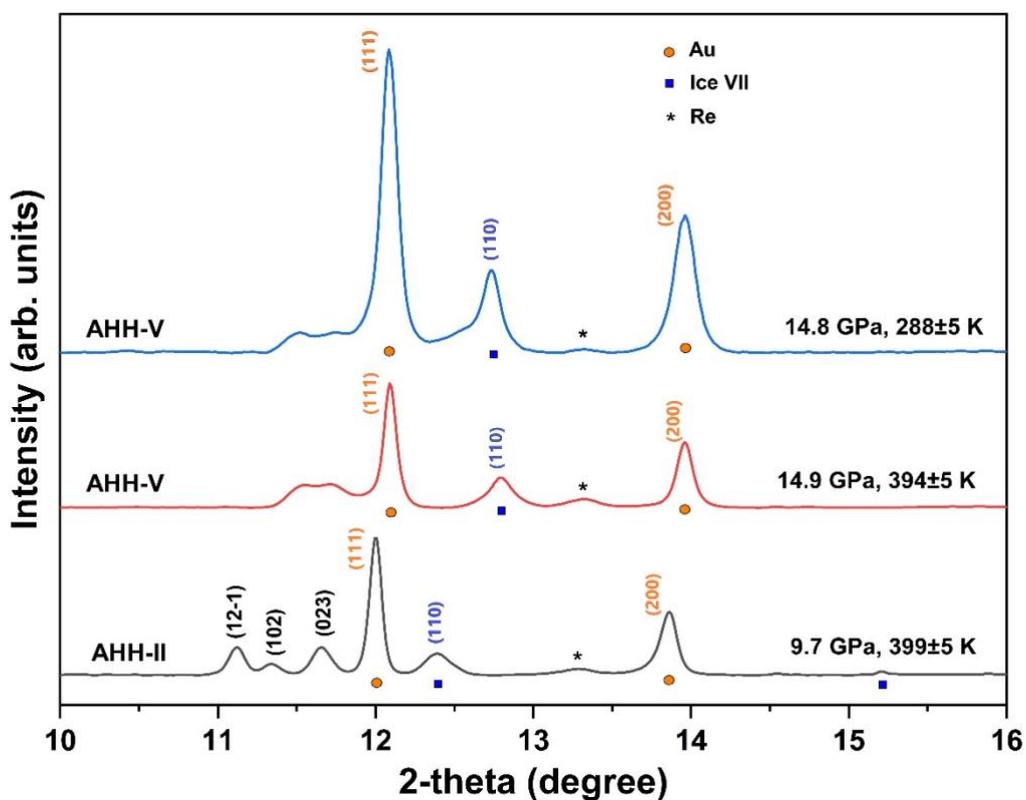

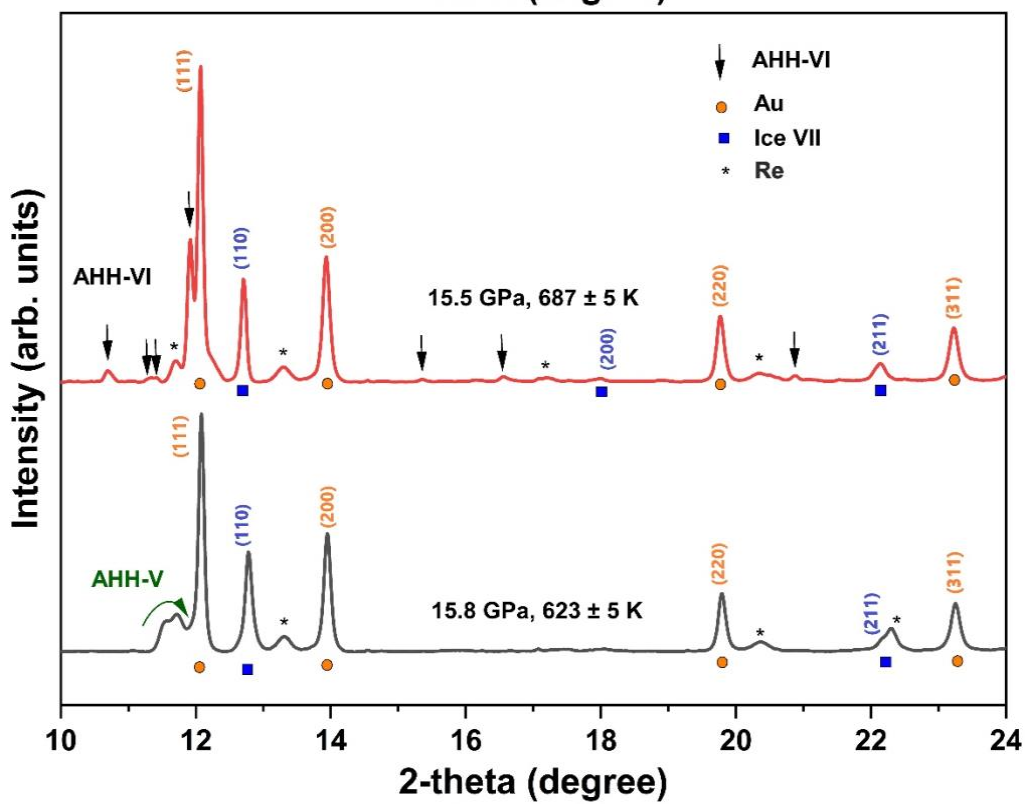

**Fig. 1. Polymorphic phase transitions of AHH in the binary AHH – ice VII system.** Selected X-ray diffraction patterns (incident X-ray energy = 25.6 keV) showing the transition from AHH-II to AHH-V (a) and from AHH-V to AHH-VI (b). Compression of monoclinic AHH-II between 13- 20 GPa (< 650 K) results in the formation of the AHH-V phase, which is characterized by the loss of the (102) reflection. The phase is preserved upon quenching to room temperature (a). Upon heating above 650 K, AHH-V transforms into a previously unidentified polymorph: the two major reflections of the AHH-V phase (green curly arrow) merge to become one strong reflection, while additional weak reflections (black arrows) appear, indicating that AHH-VI has lower symmetry than the AHH-V phase (b). Plausible symmetries for the AHH-V and AHH-VI phases are discussed in Text S1 of the Supplementary Information.

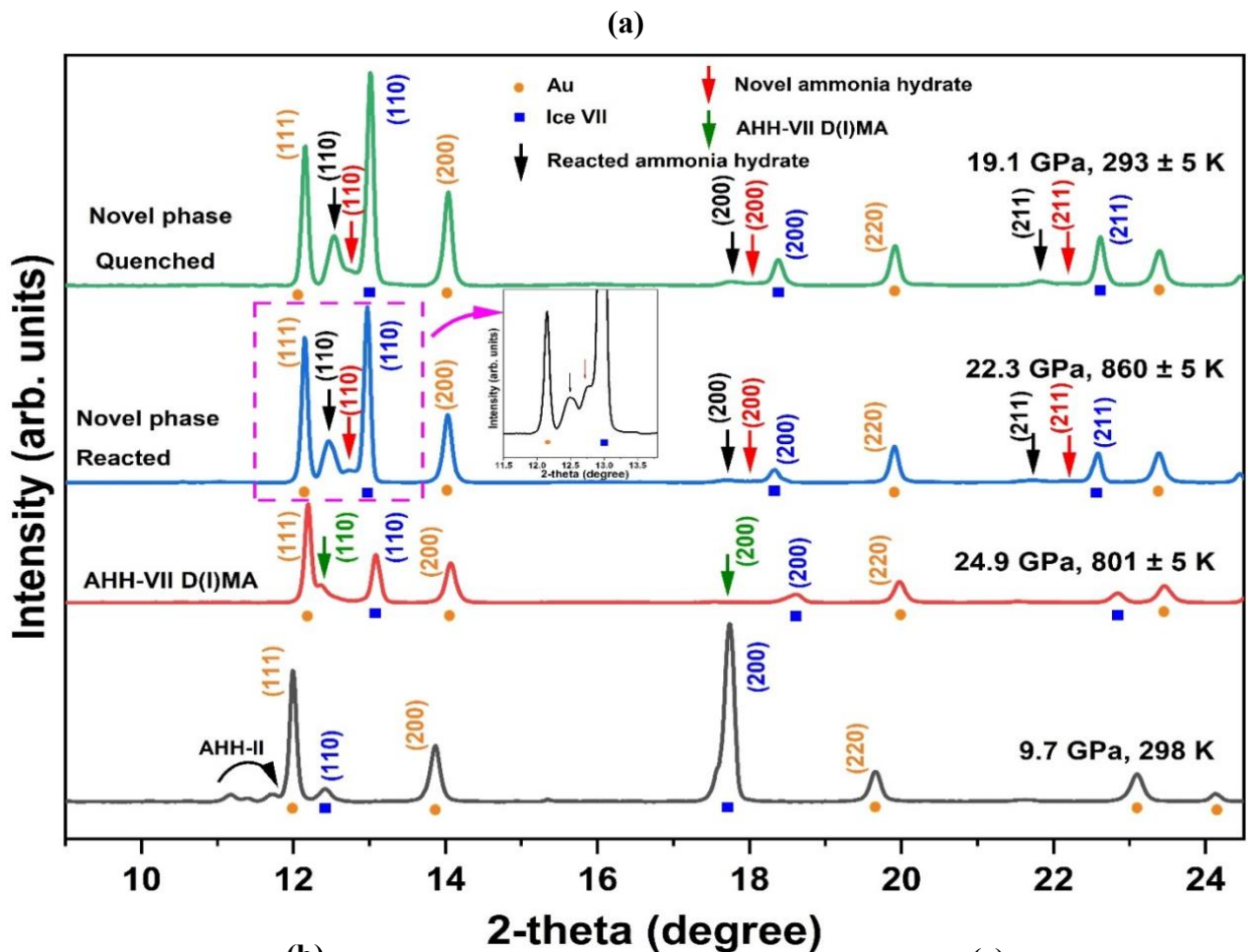
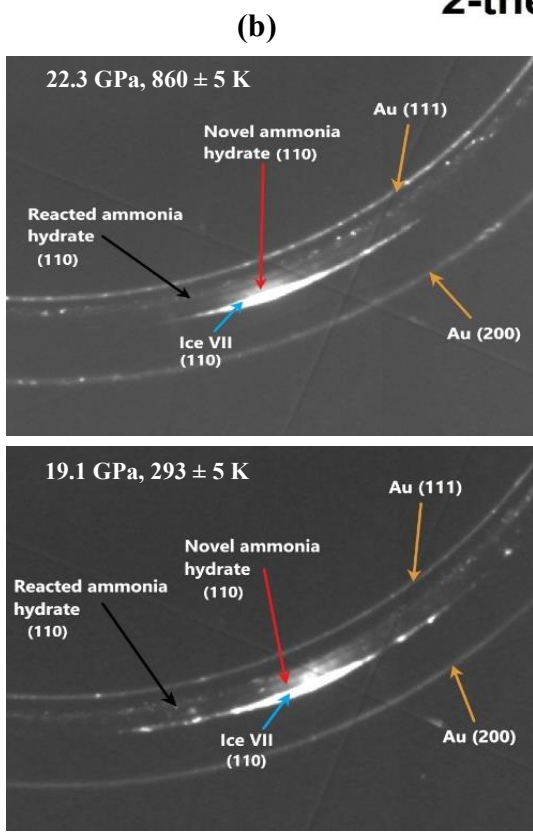
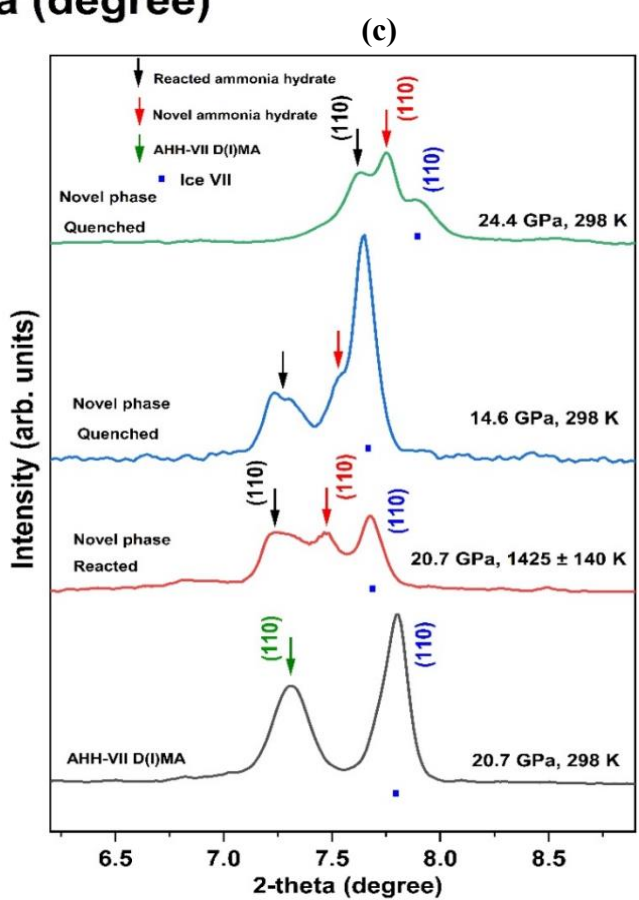

**Fig.2. X-ray diffraction patterns showing the reaction between AHH and ice VII in different runs.** Resistive heated dynamic DAC experiments **(a, b)**; Laser heated DAC experiments **(c)**. AHH-II (monoclinic) transforms into bcc AHH-VII D(I)MA phase upon compression below 800 K in the presence of ice VII, as observed for the pure AHH system (48). The reaction between AHH-VII D(I)MA phase and ice VII results in the formation of a new phase assembly that includes a novel ammonia hydrate (inset in **a** for details) coexisting with a reacted ammonia hydrate (later identified as ADH-DMA) and excess ice VII **(a, c)**. 2D diffraction images recorded at high temperature (top) and after quenching (bottom) **(b)**. The novel ammonia hydrate phase can be quenched to room temperature **(a-c)** and it is stable upon further compression up to at least 25 GPa at 298 K **(c)**. X-ray diffraction patterns are arranged in chronological order from bottom to top. Error in pressure in the RHdDAC experiments **(a)** is 10% relative at high temperature; cold pressures are reported in **(c)** and the error does not exceed 1 GPa. Note that the differences in 2θ position of the main reflections is due to the different incident energy in the experiment, i.e. 25.6 keV (RHdDAC) vs. 42.7 keV (LHDAC). Color-coded values in brackets represent the (hkl) indexation for the corresponding phases.

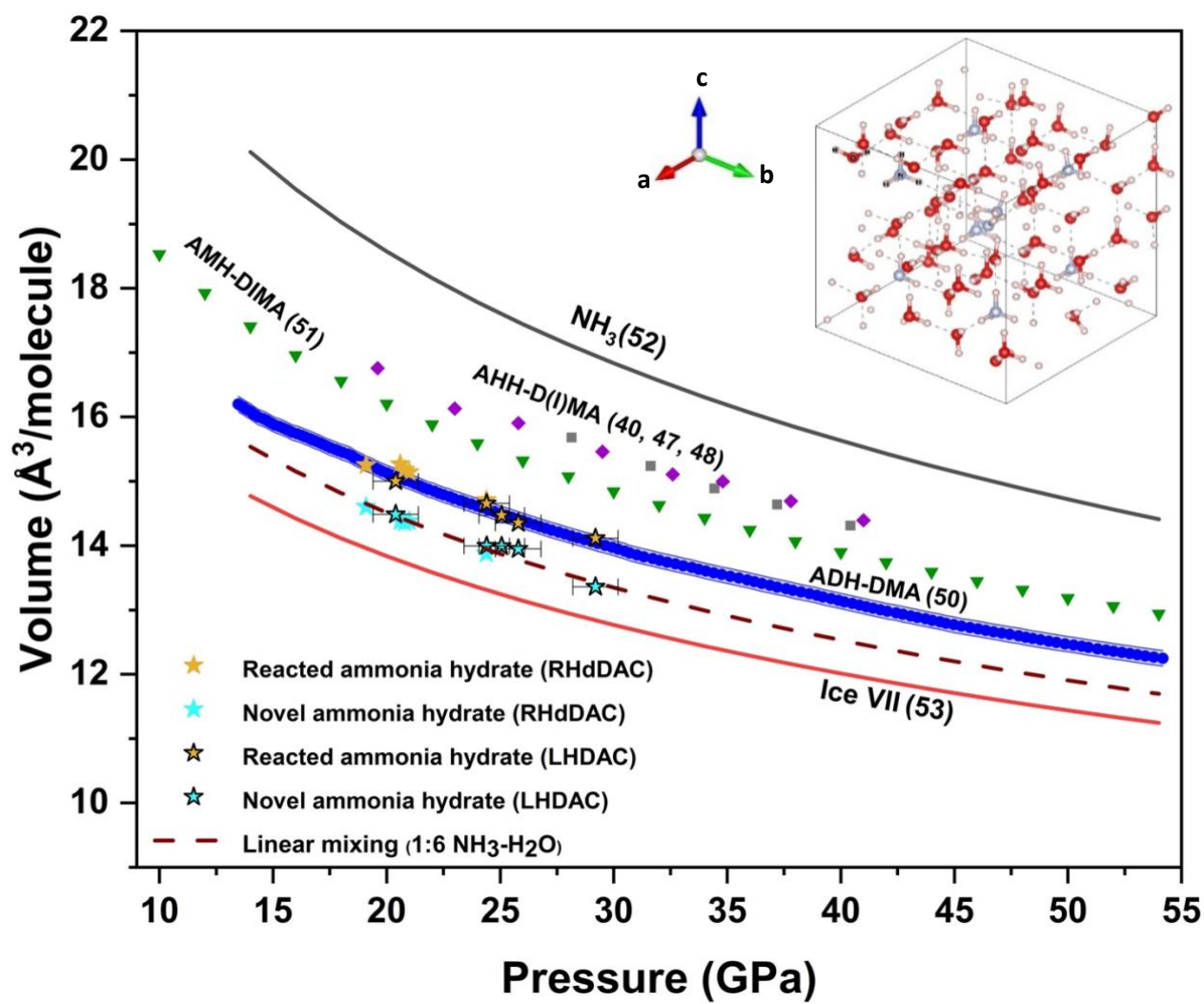

**Fig.3. Stoichiometry and structure of the novel ultra-water rich ammonia hydrate phase NH$_3$.6H$_2$O.** P-V compression curves for ammonia hydrates in the NH$_3$-H$_2$O system and *ab initio* calculated structure of NH$_3$.6H$_2$O at 20 GPa and 0 K (inset). Volumes (per molecule) retrieved after temperature quench for the 'reacted' and novel ammonia hydrate formed in the AHH-, AMH-, and ADH-ice VII binary systems are compared to volumes from other DMA/DIMA phases (40, 47, 48, 50, 51) and end-members, NH$_3$ (52) and H$_2$O (53), in the NH$_3$-H$_2$O system. The volumes of the 'reacted' phase (yellow stars) are in very good agreement with those for ADH-DMA (50), while those of the novel phase are consistent with those of an ideal NH$_3$.6H$_2$O hydrate assuming linear mixing in volume between NH$_3$ and ice VII (dashed line). Errors in volume and pressure in the RHdDAC experiments are smaller than the symbol size. Inset: Red (blue and pink) spheres denote O (N and H) atoms. Hydrogen bonds are shown by dashed black lines. The solid black line depicts the unit cell.

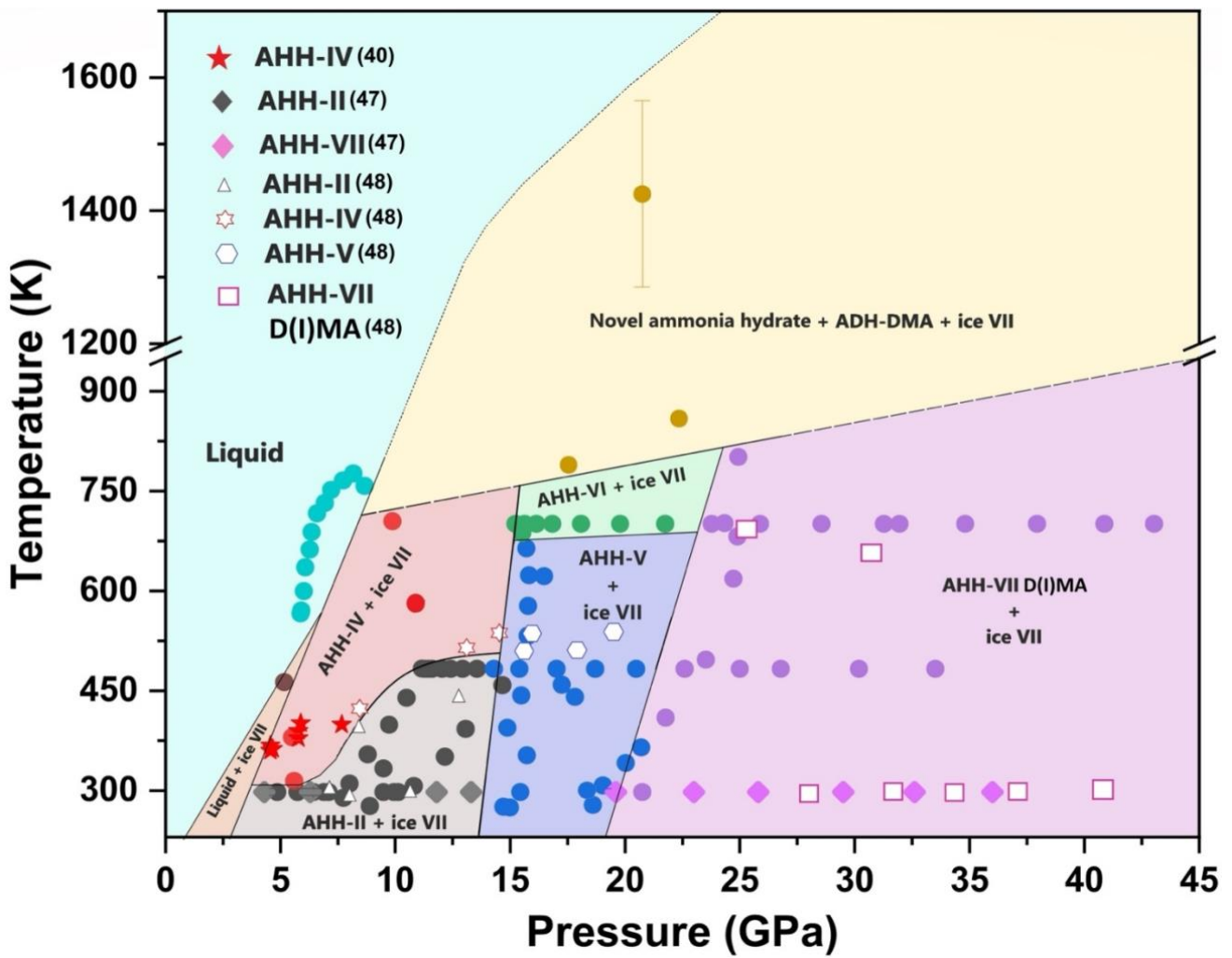

**Fig.4. Phase diagram for the AHH – H$_2$O ice VII binary system up to 45 GPa and 1500 K.** A total of 5 polymorphs are identified in the AHH system in the presence of excess H$_2$O ice VII. Data from RHdDAC (< 900 K) and LHDAC experiments in this study are represented by circles. Other symbols denote literature data for the pure AHH system (40, 47, 48). The liquidus curve is taken from Refs (39, 48) and extrapolated to higher pressure-temperature conditions (dotted line). The stability field of ice VII + liquid is taken from Refs (39, 40). Above 750 K, the reaction between AHH and ice VII results in a novel phase assembly dominated by H$_2$O-rich ammonia hydrates, ADH-DMA and a novel NH$_3$.6H$_2$O phase, and excess ice VII. This assembly also appears upon reaction between ice VII and the other stable ammonia hydrates, ADH and AMH, and is stable up to at least 30 GPa and 1600 K (see Supplementary Information). Unless indicated, errors in temperature are ±5 K; errors in pressure are 6% below 600 K and 10% above this temperature. Details on the specific P-T path in each experimental run conducted to generate the present dataset are provided in *SI Appendix,* Fig. S1.

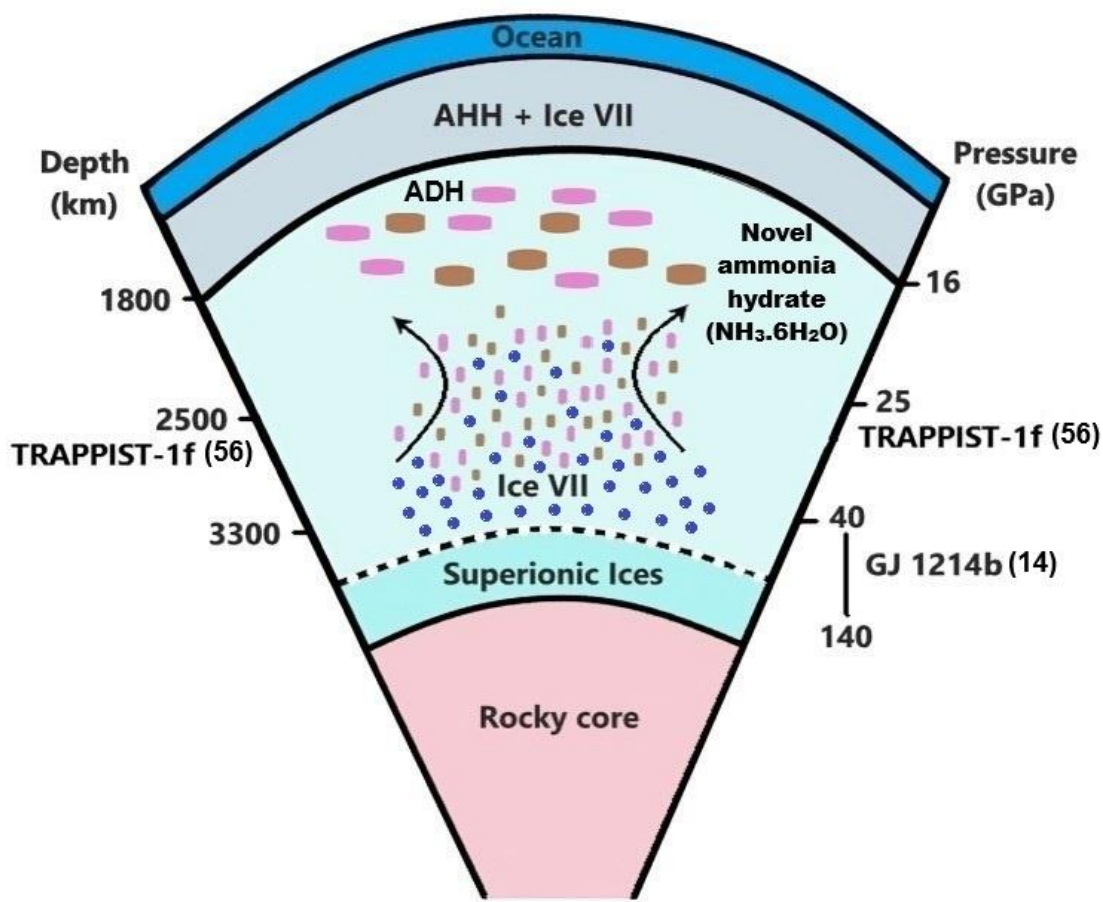

**Fig.5. Buoyancy-driven layering in the icy mantle of water-rich exoplanets.** The assembly ADH-DMA + NH$_3$.6H$_2$O + H$_2$O ice VII will be stabilized below 1800 km depth in the interior of TRAPPIST-1f like icy bodies as long as temperatures above 750-800 K are reached. The buoyancy contrast between the water-rich ammonia hydrates, ADH-DMA and NH$_3$.6H$_2$O, and H$_2$O ice (Fig. 3) will promote the formation of ammonia-bearing upwellings within the icy mantle layer, leading to phase separation and compositional layering. Note that ammonia concentration gradients may be plausible in turn within the segregated layer due to the density contrast between the ADH and NH$_3$.6H$_2$O phases (Fig. 3). Indicated pressures correspond to the reaction pressure at 750 K between AHH-D(I)MA and H$_2$O ice VII (Fig. 3) and to the pressure at ice/rock interface of water-rich exoplanets such as TRAPPIST-1f (56) and GJ 1214b (14), where 40-140 GPa indicates the range of pressures estimated depending on the bulk H$_2$O content assumed in the interior model). Note: Figure is not drawn to scale.

# Supplementary Information for

## Evidence for ultra-water-rich ammonia hydrates stabilized in icy exoplanetary mantles

A. Mondal[1], K. Mohrbach[1,2], T. Fedotenko[2], M. Bethkenhagen[3], H.-P. Liermann[2], C. Sanchez-Valle[1] [*]

[1] Institut für Mineralogie, Universität Münster, 48149 Münster, Germany

[2] Deutsches Elektronen-Synchrotron DESY, Notkestrasse 85, 22607 Hamburg, Germany

[3] Ecole Polytechnique-LULI, Av. Fresnel, 91120 Palaiseau, France

[*] Corresponding author: sanchezm@uni-muenster.de

**This pdf file includes:**

Text S1.

Figure S1 to S13.

Tables S1 and S2.

**Supplementary text.**

**Text S1. Plausible structures of AHH-V and AHH-VI polymorphs.** Appropriate structure solutions for the AHH-V phase were not possible due to the low number of visible reflections in the XRD patterns (Fig. S3). In an attempt to identify the structure of AHH-V, we compared the experimental XRD patterns to those simulated for plausible quasi-bcc structures ($P\bar{1}$, $A2/m$, $Cm$) (41) consistent with the high symmetry observed (Fig. S5). Nevertheless, we do not find a reasonable match with any of the predicted structures and hence, the structure of AHH-V in the binary system is yet to be resolved.

The transition from AHH-V to the previously unidentified AHH-VI phase is characterized by the merge of the two strong reflections of AHH-V at 2θ of 11.55° and 11.72° into one strong reflection at 2θ of 11.92° and by the appearance of additional weak reflections at lower and higher 2θ angles (Fig. 1b). We utilized the DICVOL algorithm (86) to search for a possible unit-cell solution for the AHH-VI phase and predict a monoclinic crystal structure (M = 103.6, F = 79.4) based on a total of 7 sample reflections. Further, we employed the CHEKCELL software (87) to determine the most plausible space group solution of the proposed monoclinic structure. The best-fit solution is consistent with a $P2_1/m$ space group with unit-cell parameters: a = 5.4149 ± 0.0028 Å, b = 2.7585 ± 0.0009 Å, c = 5.2075 ± 0.0023 Å and β = 93.83° ± 0.18°. An alternative space group solution could be a quasi-bcc structure $C2/m$ (41) with unit-cell parameters: a = 5.4135 ± 0.0014 Å, b = 2.7600 ± 0.0006 Å, c = 5.2086 ± 0.0014 Å and β = 93.82° ± 0.09°, although a lower number of reflections could be matched accurately with this solution.

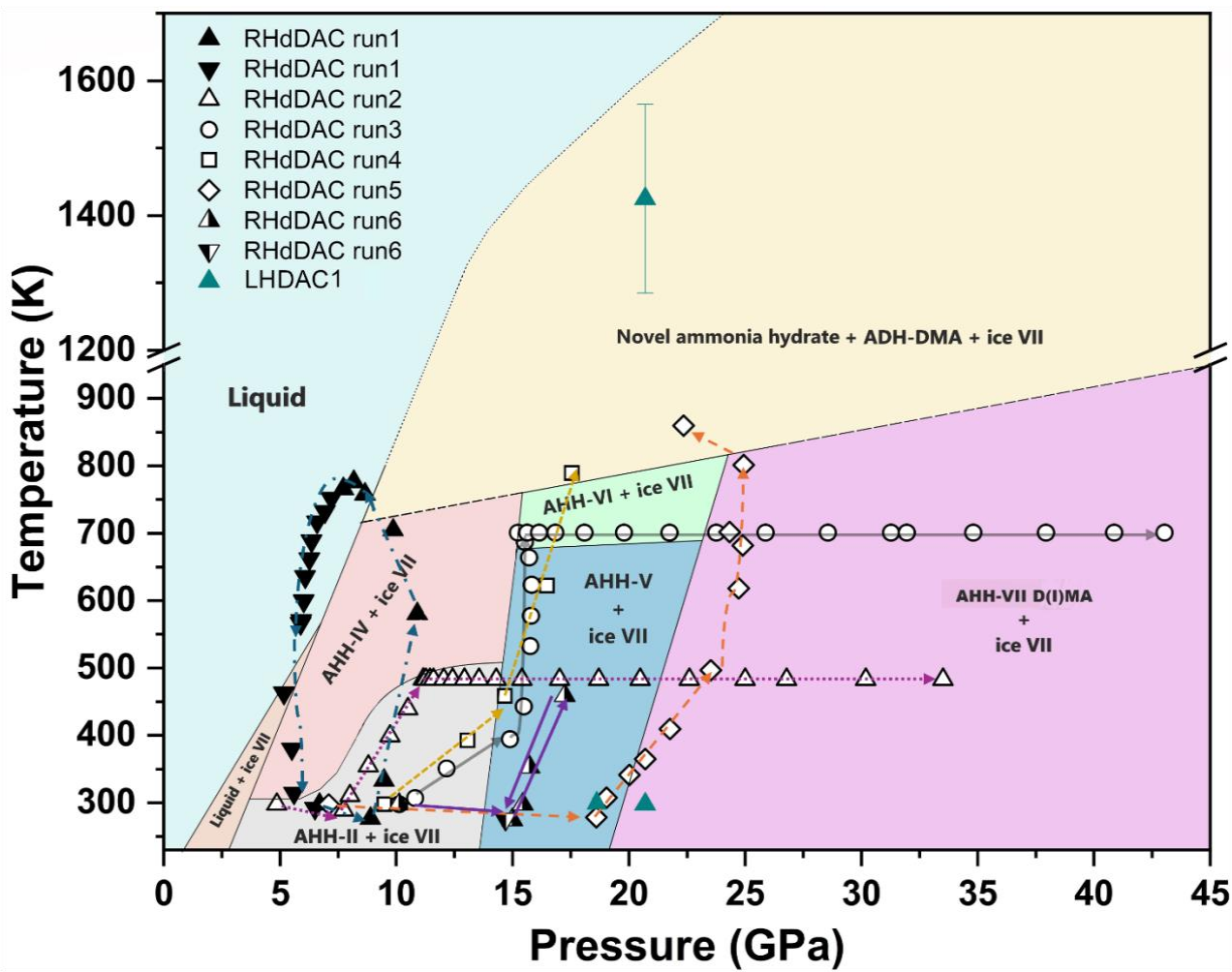

**Fig. S1. Pressure-temperature coverage of the data.** Phase boundaries for the polymorphic transformations (solid lines) are those identified from the XRD analysis in the present study. The liquidus curve is taken from previous studies (39, 48) and extrapolated to higher pressure-temperature (P-T) conditions (dotted line). The stability field of liquid + ice VII is taken from previous reports (39, 40). Data were collected upon increasing temperature unless indicated by the pointing down symbols (RHdDAC run1 and run6). The arrows are guides to the eyes to indicate the P-T path followed in each run. Data recorded along isotherms of 483 K (empty triangles) and 701 K (empty circles) were recorded upon continuous compression in the RHdDAC. Unless indicated, errors in temperature are ±5 K; errors in pressure are 0.6% relative below 600 K and 1% above this temperature for the RHdDAC experiments; cold pressure error in the LHDAC data is ±1 GPa (see details in the Methods). This applies for all the reported data in the Supplementary Information (Figs. S2 to S9).

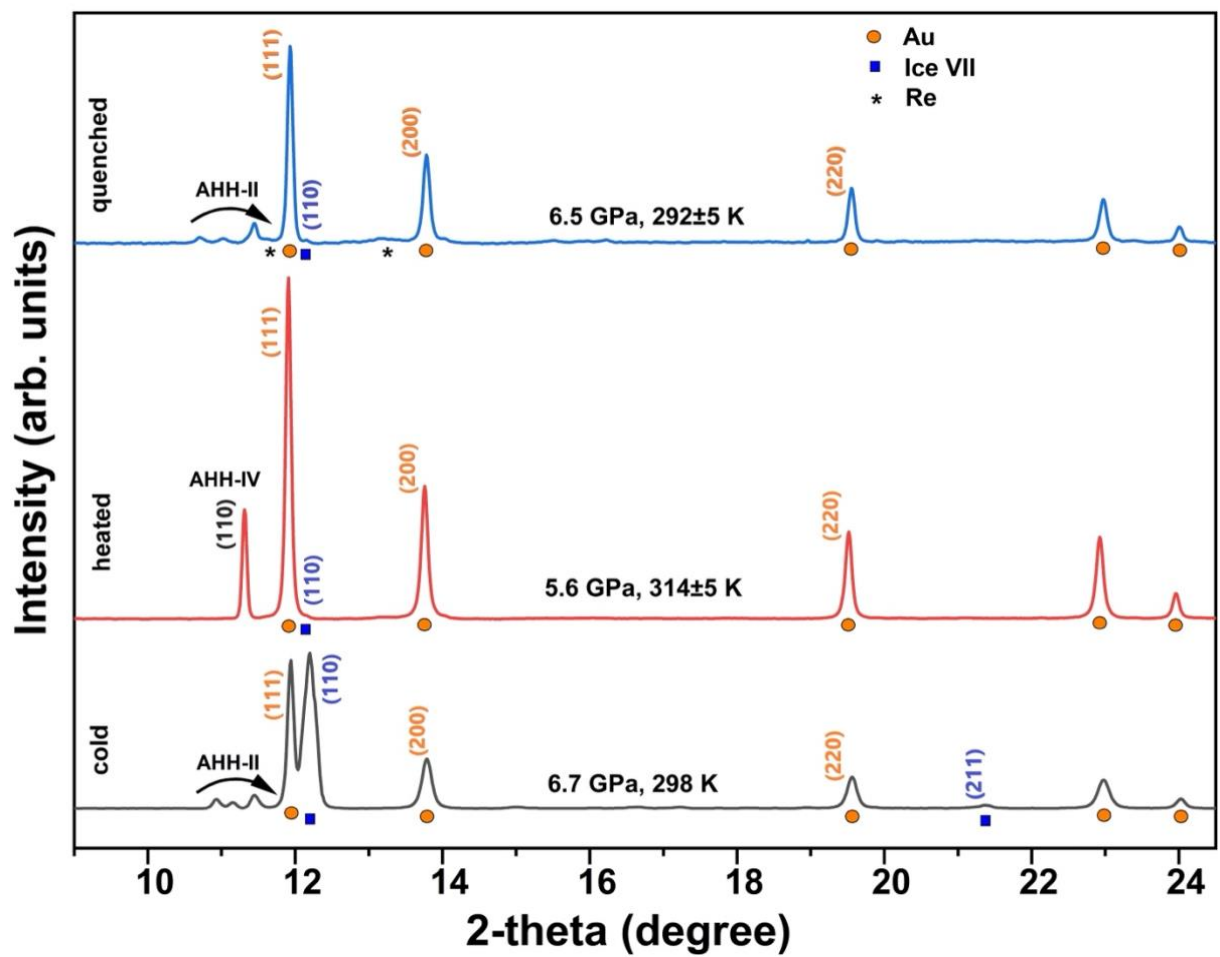

**Fig. S2. Reversible polymorphic phase transition from AHH-II to AHH-IV upon heating.** Selected XRD patterns (incident x-ray energy = 25.6 keV) displaying the AHH-II to AHH-IV transition upon heating. The three characteristic strong reflections of the AHH-II phase merged to become one strong reflection in the bcc AHH-IV phase. The phase reverts into AHH-II upon quenching to room temperature as indicated by the similarities between the diffraction patterns, hence suggesting a reversible transition.

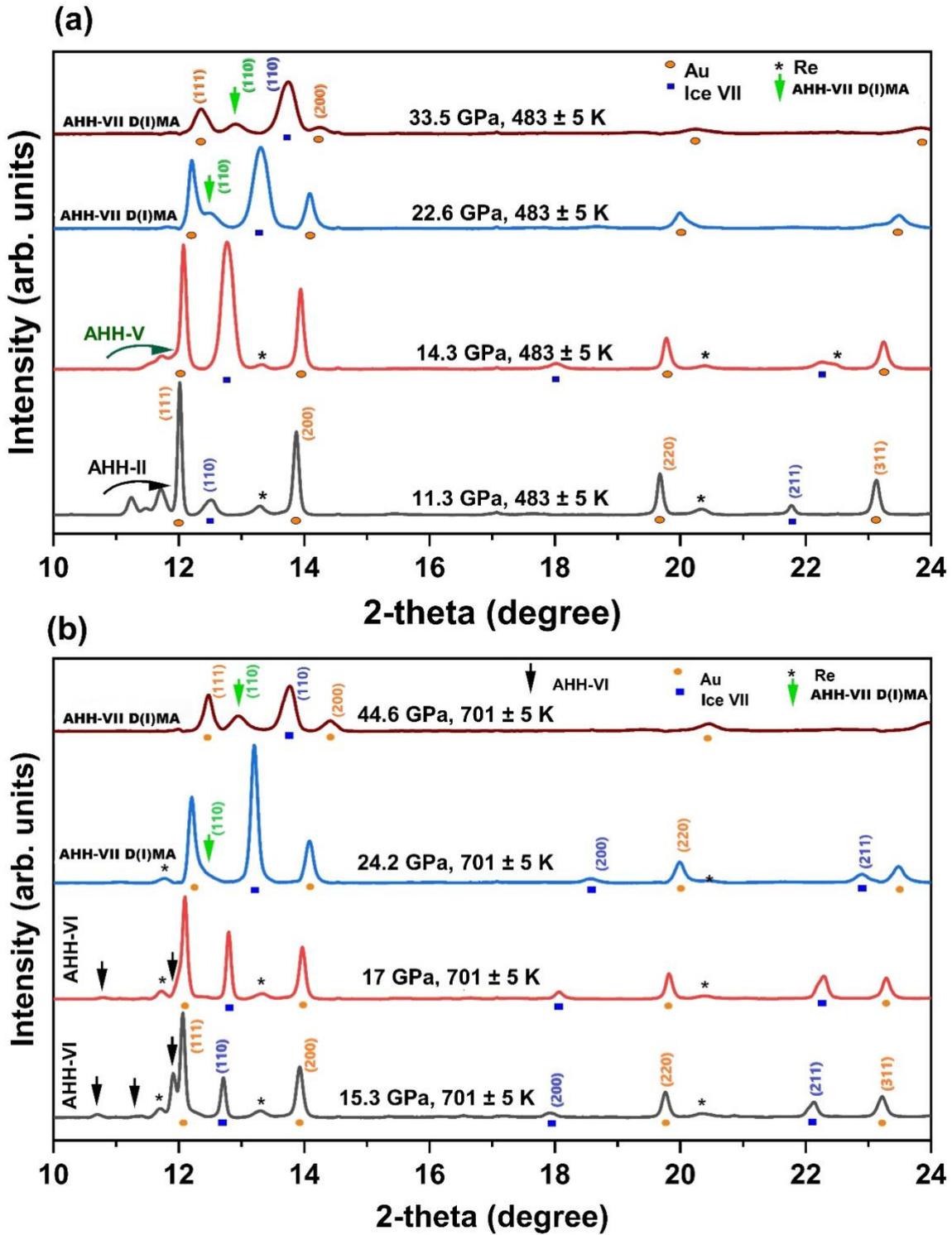

**Fig. S3. Ramp compressions at constant temperature in the RHdDAC.** Selected diffraction patterns (incident x-ray energy = 25.6 keV) from the ramp compressions preformed at constant temperature of (a) 483 ± 5 K, displaying the transformation from AHH-II to V and VII D(I)MA (7) upon increase in pressure, and (b) 701 ± 5 K, showing the transformation from AHH-VI to AHH-VII D(I)MA (47) phase, respectively.

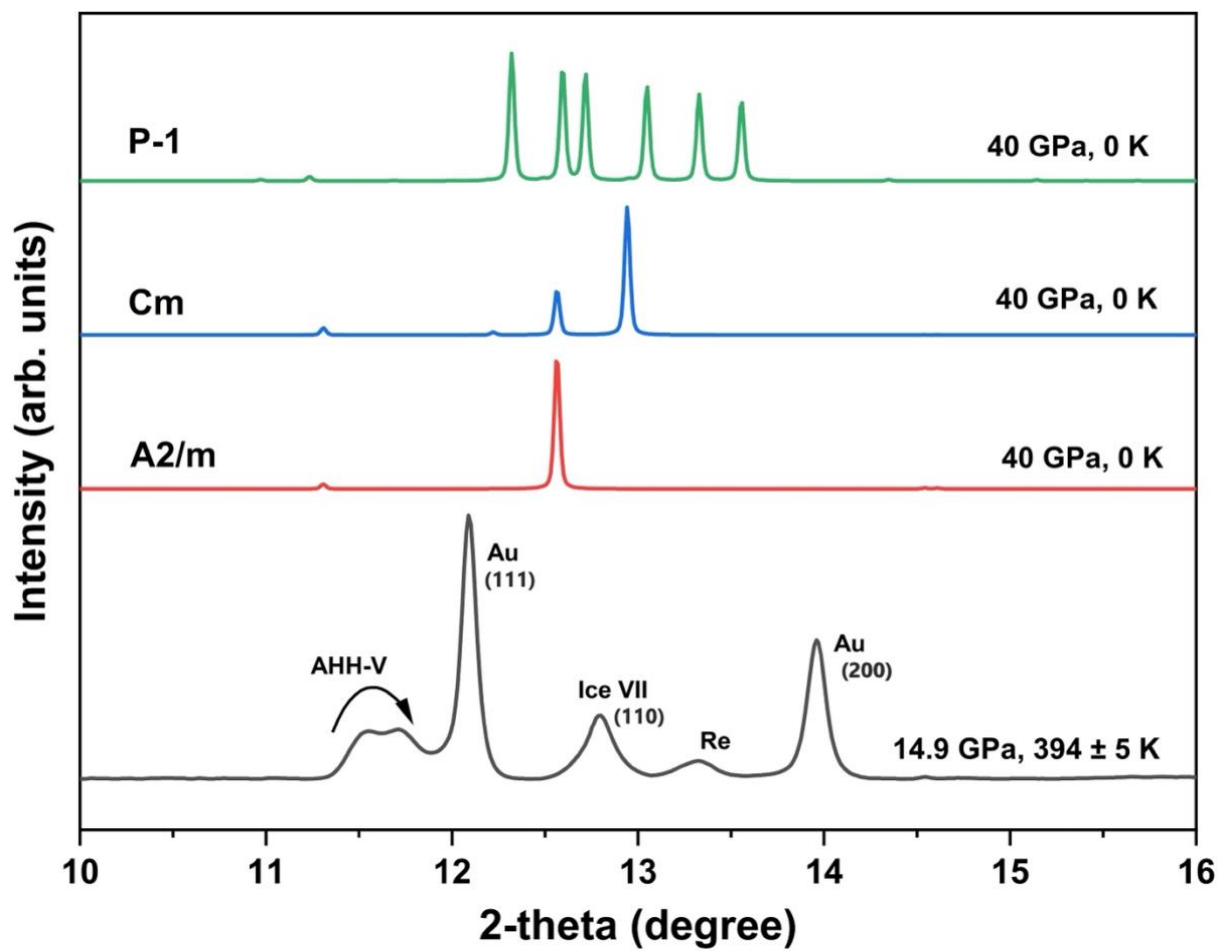

**Fig. S4. Crystal structure of AHH-V phase.** Comparison between the x-ray diffraction pattern (incident x-ray energy = 25.6 keV) of the AHH-V phase and the quasi-bcc structures predicted by theoretical simulations (41). Regardless of the difference in pressure, none of the predicted structures match the experimental patterns of the AHH-V phase.

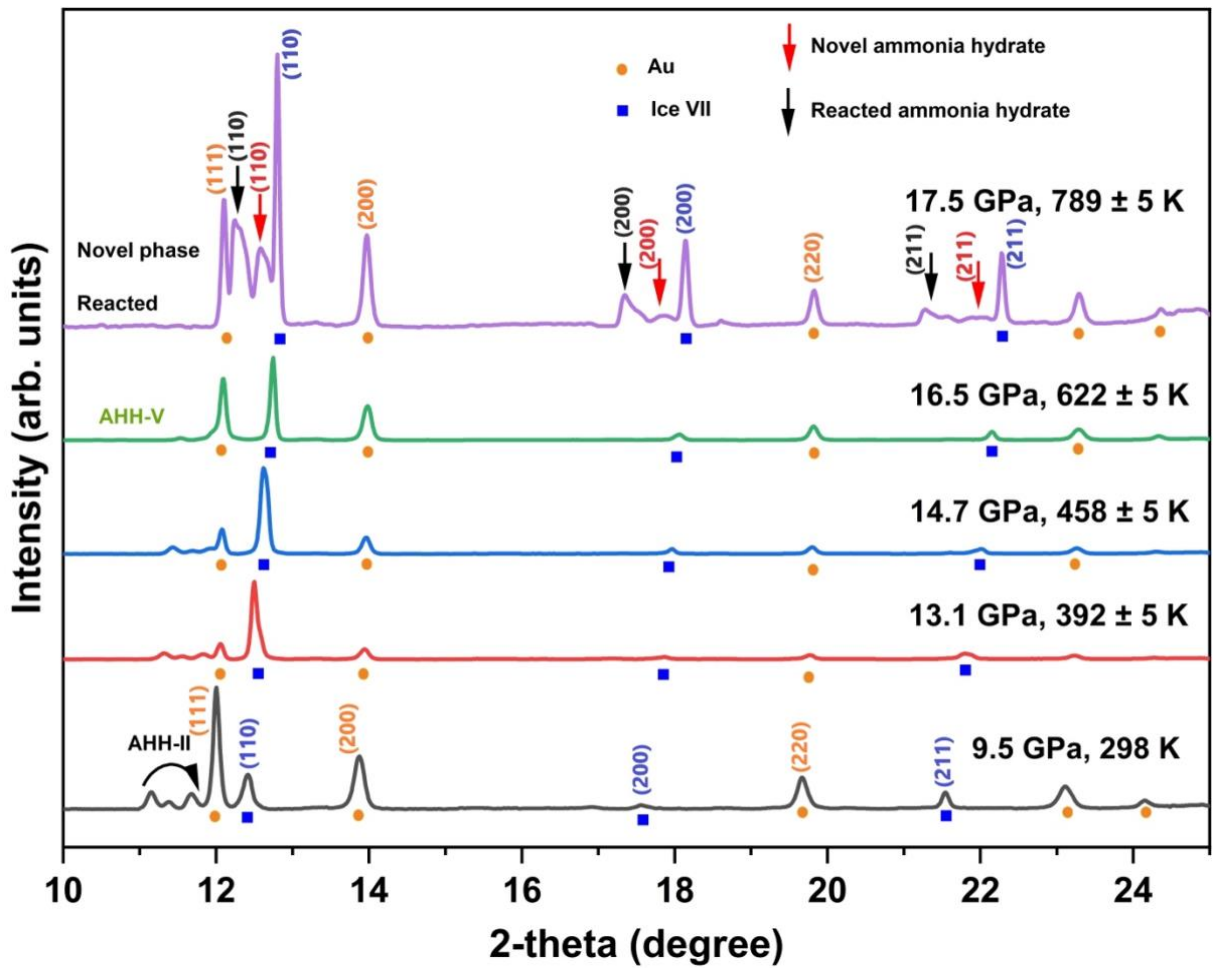

**Fig. S5. X-ray diffraction patterns showing the reaction between AHH and ice VII.** XRD patterns (incident x-ray energy = 25.6 keV) stacked in chronological order from bottom to top collected in RHdDAC (run4, Fig. S1). At the highest temperature, 789 K, the reaction between AHH and $H_2O$ ice VII results into a novel phase assembly consisting of the reacted (black arrow) and novel ammonia hydrate (red arrow) phases, coexisting with excess ice VII.

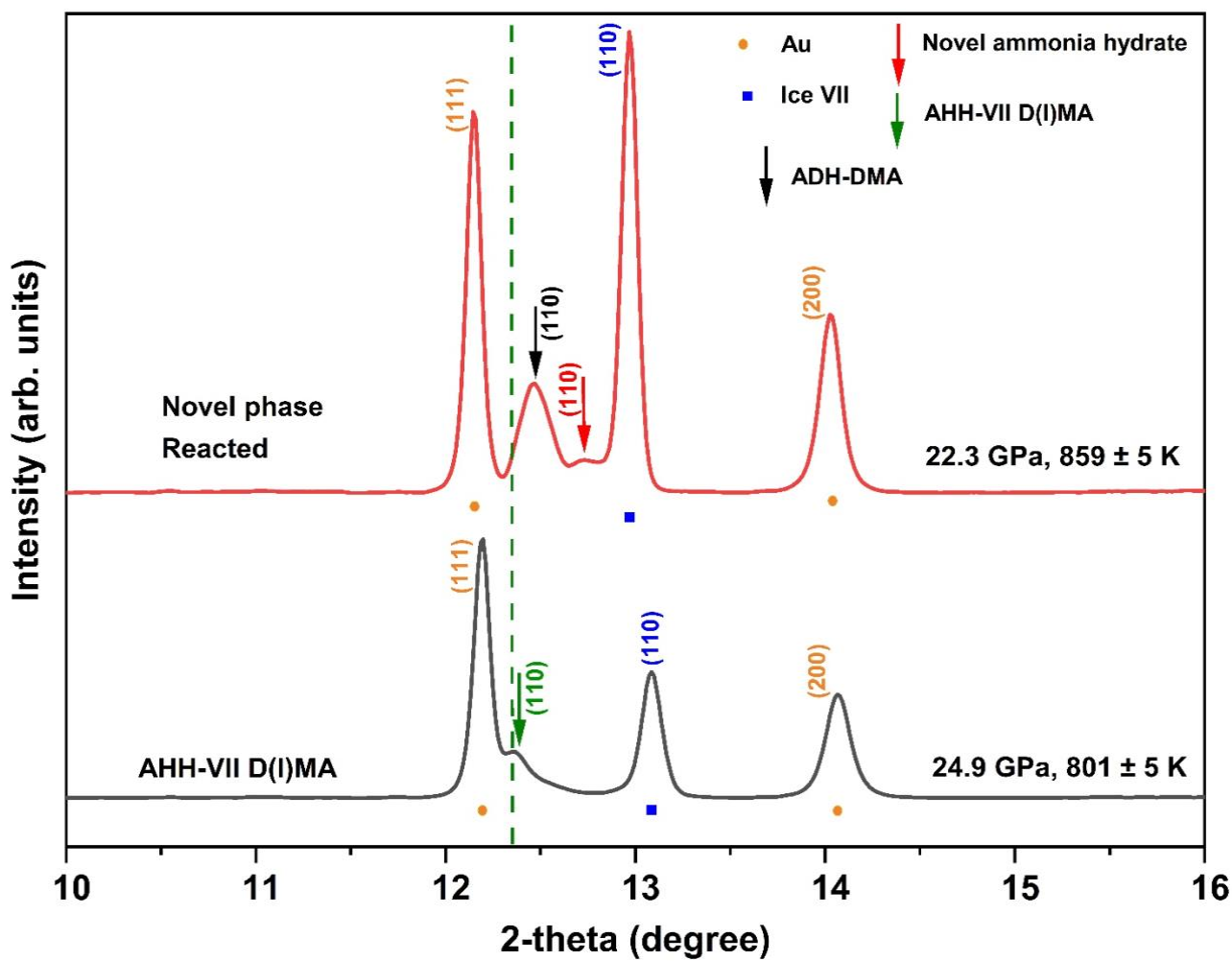

**Fig. S6. Changes in the XRD pattern upon reaction between AHH and ice VII.** X-ray diffraction patterns (incident x-ray energy = 25.6 keV) displaying the differences before and after the reaction between AHH-VII D(I)MA and $H_2O$ ice VII. The reflection of the AHH-VII D(I)MA phase (green arrow) is replaced by two new reflections that corresponds to the ADH-DMA (black arrow) and to the novel ammonia hydrate phase (red arrow).

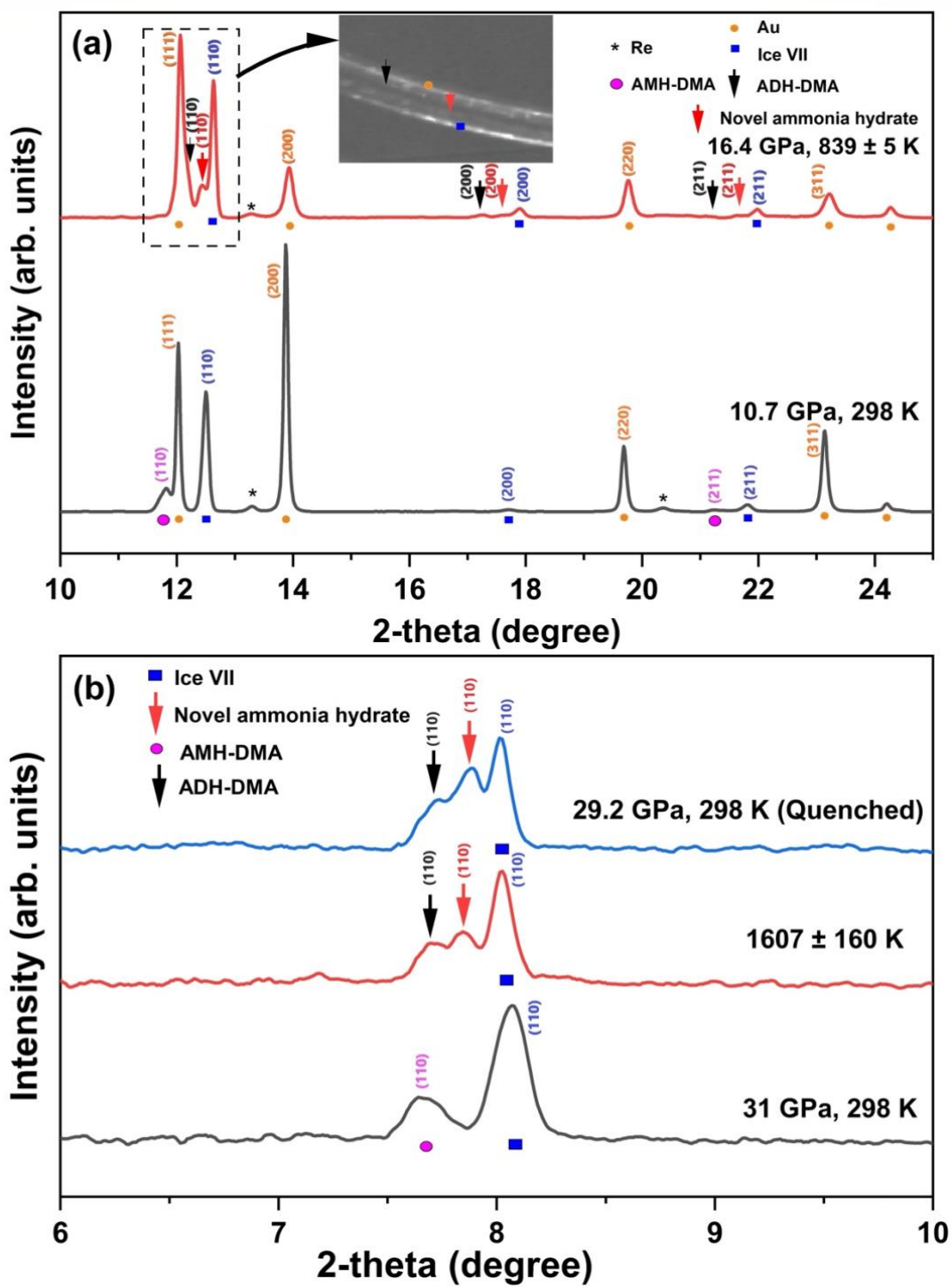

**Fig. S7. XRD patterns showing the reaction between AMH and ice VII in different run conditions.** RHdDAC experiments, incident x-ray energy = 25.6 keV (a); LHDAC experiments, incident x-ray energy= 42.7 keV (b). The reaction is characterized by the appearance of new reflections ascribed to ADH-DMA and to the novel $H_2O$-rich ammonia hydrate $NH_3 \cdot 6H_2O$ (see main text for additional discussion). The new reflections are also observed in the raw XRD image (inset). The reacted phase assembly can furthermore be quenched to room temperature condition, as observed for the AHH-ice VII system.

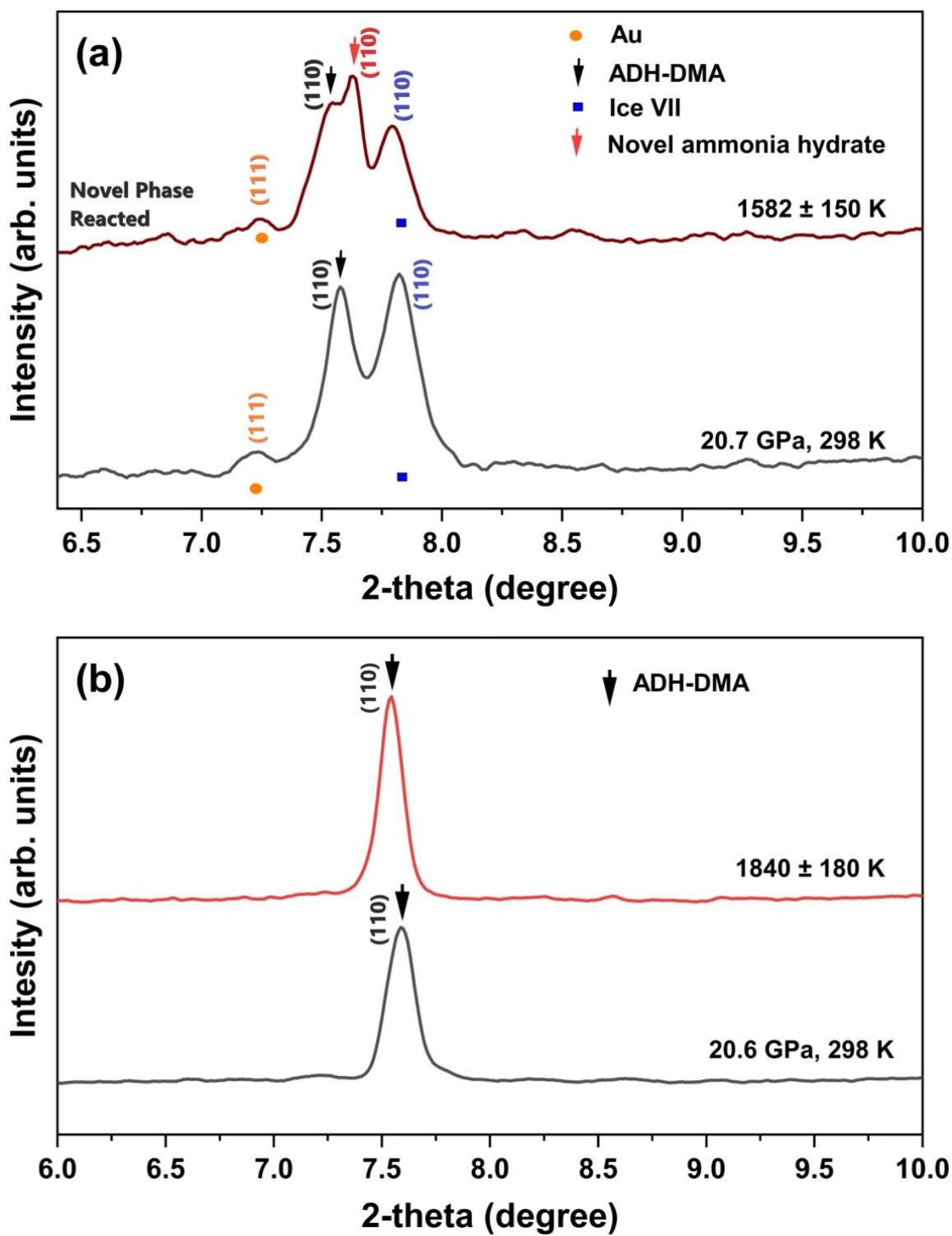

**Fig. S8. XRD patterns recorded in the ADH-ice VII and ADH samples during laser-heating.** The appearance of the ADH-DMA + $NH_3 \cdot xH_2O$ (with x= 6±1) + ice VII assembly is observed above 20 GPa and 1582 ± 150 K (a). In the absence of excess ice VII in the system, the ADH-DMA phase remains stable up to at least 1840 ± 180 K without breakdown or dehydration as reported below 10 GPa (40) (b).

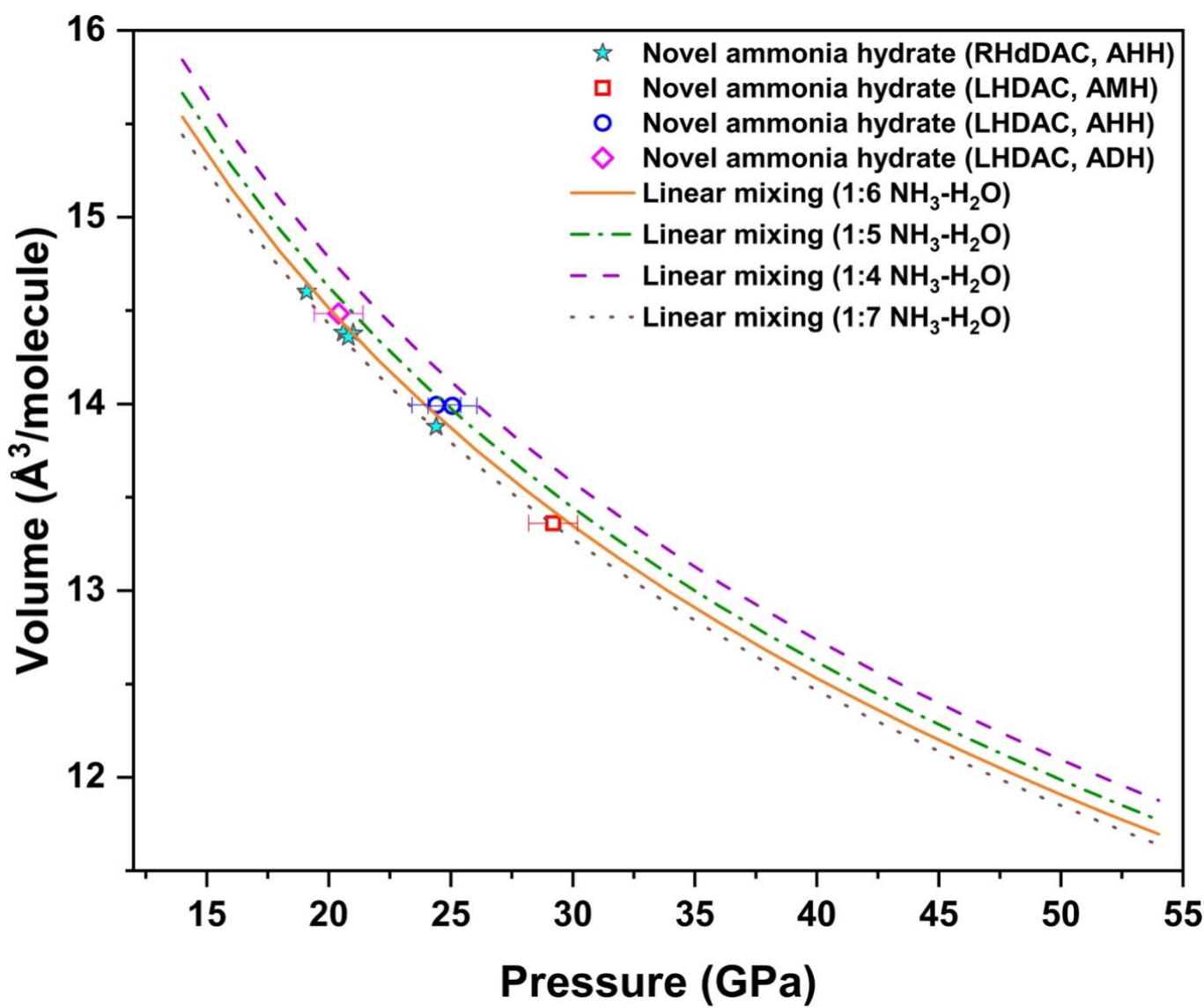

**Fig. S9. P-V data of the novel ammonia hydrate phase compared with compression curves of hypothetical NH₃.xH₂O hydrates, with x = 4, 5, 6 and 7.** Experimental volumes for the novel hydrate were retrieved after temperature quenching of the reacted assembly and further compression at room temperature in the AHH-, AMH- and ADH- ice VII binary systems in both RHdDAC and LHDAC experiments (Table S2). The compression curves for the hypothetical hydrates were calculated under the assumption of linear mixing in volume between NH₃ and ice VII (54). The data is better explained by the 1:6 molar ratio (i.e. NH₃.6H₂O stoichiometry) considering the mutual uncertainties in pressures and volumes. Errors in volume and pressure in the RHdDAC experiments are smaller than the symbol size.

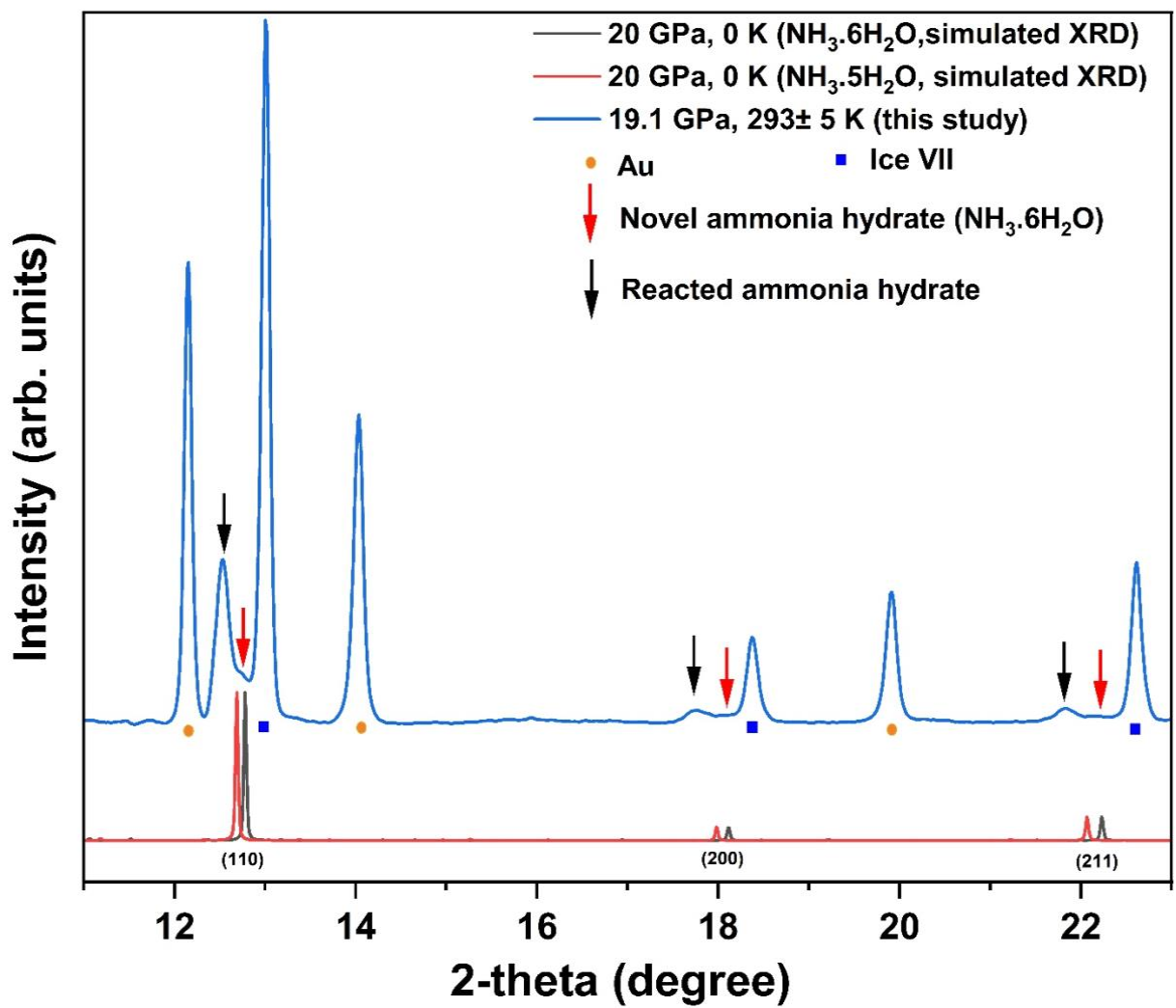

**Fig. S10. Experimental XRD pattern of the novel ammonia hydrate NH$_3$.6H$_2$O and comparison with simulated patterns of NH$_3$·xH$_2$O, where x = 5 and 6.** The experimental pattern was collected in the RHdDAC (run5, incident x-ray energy = 25.6 keV) after quench.

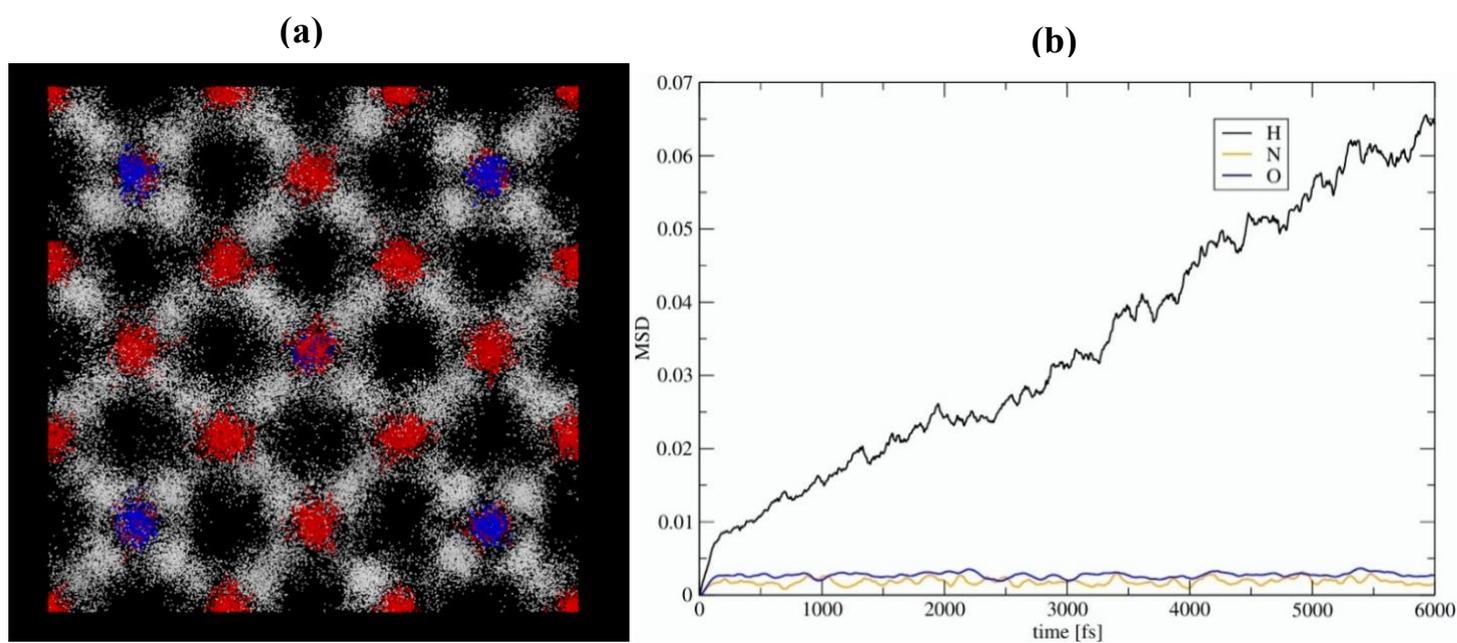

**Fig. S11. Structure and stability of NH$_3$.5H$_2$O ammonia hydrate at 19.5 GPa and 900 K.** *Ab initio* molecular dynamic (AIMD) simulations with the full proton trajectory over 6000 fs run shown as white dots (a). The mean square displacement (MSD, in Å) curves show that the heavy ions (N and O) remain in the bcc lattice position, hence supporting the stability of the structure at high pressure-high temperature conditions (b).

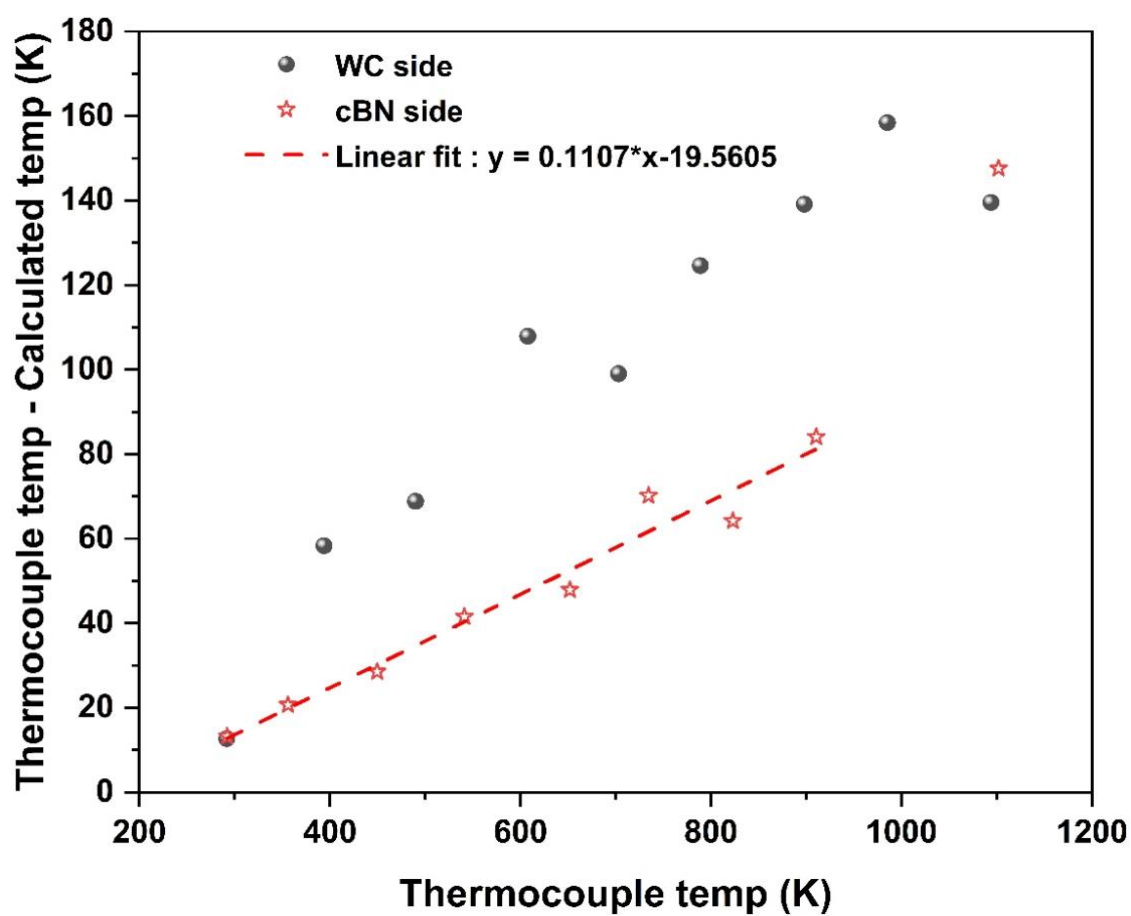

**Fig. S12. Temperature calibration in the RHdDAC.** Temperature difference between the thermocouple and the sample temperature (Au powder) plotted against the temperature of the thermocouple. Black and red symbols denote data obtained for the thermocouples placed on the WC and cBN seat sides of the RHdDAC, respectively. The temperature gradient increases linearly with temperature up to 1000 K and is smaller for the thermocouple on the cBN seat side, which is a better proxy for sample temperature during the experiments. A linear fit of the cBN side thermocouple data (dashed line) provides the temperature correction that is applied to the thermocouple reading to retrieve the actual sample temperature during the run.

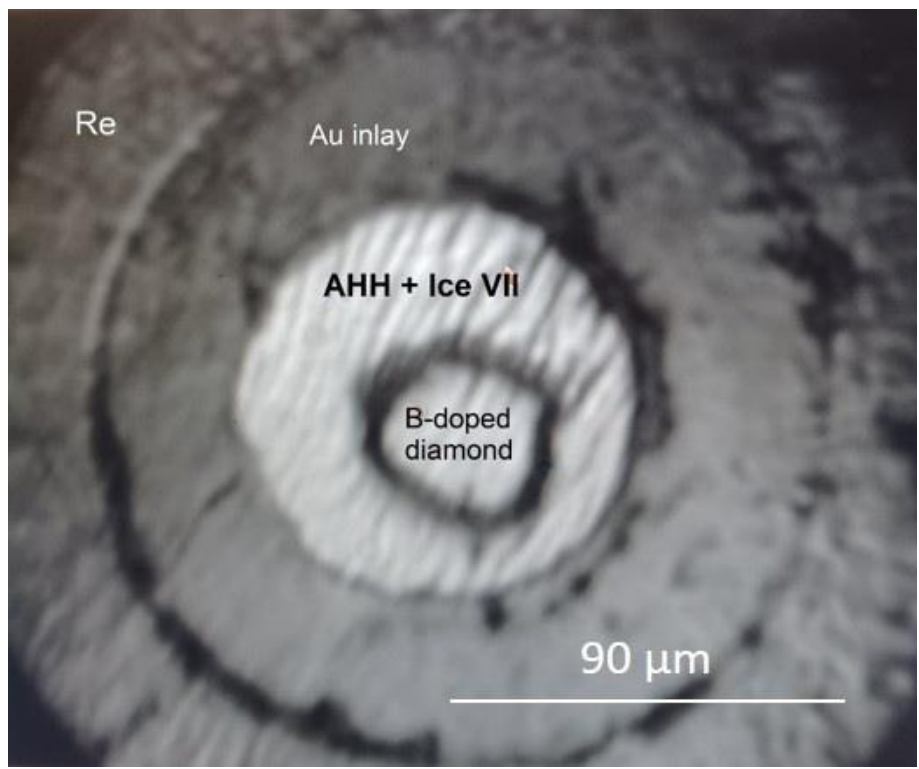

**Fig. S13. Microphotograph of the sample chamber arrangement for LHDAC experiments.** A gold (Au) inlay prevents undesired chemical reactions between the Re gasket and the sample. A ~10 μm thick B-doped diamond disk embedded into the sample chamber serve as a laser coupler. AHH-II and ice VII assembly was synthesized upon slow compression up to at least 4 GPa of an $NH_3$-$H_2O$ solution containing 32 wt% $NH_3$ at room temperature, 298 K.

**Table S1.** Summary of experimental runs where the formation of the novel ammonia hydrate assembly 'ADH-DMA + NH$_3$.6H$_2$O + ice VII' was observed in the AHH-, AMH- and ADH + ice VII systems, and lattice parameters for ADH-DMA and NH$_3$.6H$_2$O determined at room temperature after temperature quench.

| Run | Technique[1] | Initial assembly[2] | P (GPa)[3] | T (K)[3] | P$_{quench}$ (GPa)[5] | 'Reacted' phase (ADH-DMA) | | 'Novel' phase (NH$_3$.6H$_2$O) | |
|---|---|---|---|---|---|---|---|---|---|
| | | | | | | a (Å) | V(Å$^3$/molec.) | a (Å) | V(Å$^3$/molec.) |
| **AHH + ice VII system** | | | | | | | | | |
| Run4 | RHdDAC | AHH-V + ice VII | 17.5 ± 0.2 | 789 ± 5 | N.N[6] | - | - | - | - |
| Run5 | RHdDAC | AHH-VII + ice VII | 22.3 ± 0.2 | 860 ± 5 | 19.1 ± 0.2 | 3.124 ± 0.004 | 15.250 ± 0.058 | 3.079 ± 0.004 | 14.602 ± 0.057 |
| LHDAC1 | LHDAC | AHH-VII + ice VII | 20.7 ± 1.0 | 1425 ± 140 | 14.6 ± 1.0 | N.N[7] | N.N[7] | N.N[7] | N.N[7] |
| LHDAC2 | LHDAC | AHH-VII + ice VII | 30.2 ± 1.0 | N.N[4] | N.N[6] | - | - | - | - |
| LHDAC3 | LHDAC | AHH-VII + ice VII | 18.6 ± 1.0 | N.N[4] | N.N[6] | - | - | - | - |
| **AMH + ice VII system** | | | | | | | | | |
| RHdDAC1a | RHdDAC | AMH-DMA + ice VII | 16.4 ± 0.2 | 839 ± 5 | N.N[6] | - | - | - | - |
| LHDAC1a | LHDAC | AMH-DMA + ice VII | 31 ± 1 | 1607 ± 160 | 29.2 ± 1.0 | 3.045 ± 0.004 | 14.113 ± 0.056 | 2.990 ± 0.004 | 13.360 ± 0.054 |
| **ADH + ice VII system** | | | | | | | | | |
| LHDAC1b | LHDAC | ADH-DMA + ice VII | 20.7 ± 1.0 | 1582 ± 150 | 20.4 ± 1.0 | 3.107 ± 0.004 | 15.000 ± 0.058 | 3.071 ± 0.004 | 14.485 ± 0.056 |

Notes: [1] RHdDAC - Resistive heated dynamic DAC; LHDAC – Laser Heated DAC. [2] Last phase assembly identified by XRD before the reaction was observed. [3] P-T conditions where the reacted assembly was confirmed by XRD. [4] Temperature could not be measured by spectro-radiometry (see Methods). [5] Pressure recorded at room temperature after temperature quench. [6] Experiment terminated before quenching due to DAC failure at high temperature. [7] Unit-cell parameters could not be determined due to overlap with other reflections.

**Table S2.** Lattice parameters of ADH-DMA and NH$_3$.6H$_2$O phases determined after temperature quench and further compression of the recovered assemblage at 298 K in the AHH-, AMH- and ADH-ice VII systems. The data are displayed in Fig. 3 and *SI Appendix,* Fig. S9.

| Run | Technique[1] | P (GPa) | 'Reacted' phase (ADH-DMA) | | 'Novel' phase (NH$_3$.6H$_2$O) | |
|---|---|---|---|---|---|---|
| | | | a (Å) | V(cm$^3$/molec.) | a (Å) | V(cm$^3$/molec.) |
| **AHH + ice VII system** | | | | | | |
| Run5 | RHdDAC | 19.1 ± 0.2 | 3.124 ± 0.004 | 15.250 ± 0.058 | 3.079 ± 0.004 | 14.602 ± 0.057 |
| | | 20.6 ± 0.2 | 3.125 ± 0.004 | 15.264 ± 0.059 | 3.064 ± 0.004 | 14.381 ± 0.056 |
| | | 20.8 ± 0.2 | 3.120 ± 0.004 | 15.181 ± 0.058 | 3.062 ± 0.004 | 14.358 ± 0.056 |
| | | 21.0 ± 0.2 | 3.117 ± 0.004 | 15.138 ± 0.058 | 3.064 ± 0.004 | 14.378 ± 0.056 |
| | | 24.4 ± 0.2 | 3.087 ± 0.004 | 14.705 ± 0.057 | 3.028 ± 0.004 | 13.878 ± 0.055 |
| LHDAC1 | LHDAC | 14.6 ± 1.0 | N.N[2] | N.N[2] | N.N[2] | N.N[2] |
| | | 24.4 ± 1.0 | 3.083 ± 0.004 | 14.653 ± 0.057 | 3.036 ± 0.004 | 13.995 ± 0.055 |
| | | 25.1 ± 1.0 | 3.070 ± 0.004 | 14.470 ± 0.056 | 3.036 ± 0.004 | 13.990 ± 0.055 |
| **AMH + ice VII system** | | | | | | |
| LHDAC1a | LHDAC | 29.2 ± 1.0 | 3.045 ± 0.004 | 14.113 ± 0.056 | 2.990 ± 0.004 | 13.360 ± 0.054 |
| **ADH + ice VII system** | | | | | | |
| LHDAC1b | LHDAC | 20.4 ± 1.0 | 3.107 ± 0.004 | 15.000 ± 0.058 | 3.071 ± 0.004 | 14.485 ± 0.057 |

Notes: [1] RHdDAC - Resistive heated dynamic DAC; LHDAC – Laser Heated DAC; [2] Could not be determined due to the overlap with other reflections.